\documentclass{aa}
\usepackage[varg]{txfonts}
\bibpunct{(}{)}{;}{a}{}{,}

\usepackage{amsmath,amssymb,amsfonts}
\usepackage{caption}
\usepackage{graphicx}
\usepackage{subcaption}

\usepackage{tabularx}

\usepackage{linenoaa}

\usepackage{hyperref}
\hypersetup{
    colorlinks=true,
    linkcolor=blue,
    citecolor=blue,
    filecolor=magenta,      
    urlcolor=blue,
    pdftitle={Cermenati et al., 2025},
    pdfpagemode=FullScreen,
    }

\begin{document}

\title{Excitation of the non-resonant streaming instability around sources of Ultra-High Energy Cosmic Rays}
\titlerunning{Excitation of the NRSI around sources of UHECRs}

\author{Alessandro Cermenati\inst{\ref{inst1},\ref{inst2}}\fnmsep\thanks{alessandro.cermenati@gssi.it} \and Roberto Aloisio\inst{\ref{inst1},\ref{inst2}} \and Pasquale Blasi\inst{\ref{inst1},\ref{inst2}} \and Carmelo Evoli\inst{\ref{inst1},\ref{inst2}}}
\authorrunning{Cermenati et al.}

\institute{Gran Sasso Science Institute (GSSI), Viale Francesco Crispi 7, 67100 L’Aquila, Italy \label{inst1} \and INFN-Laboratori Nazionali del Gran Sasso (LNGS), via G. Acitelli 22, 67100 Assergi (AQ), Italy \label{inst2}}

\date{\today}

\abstract {} {The interpretation of the ultra-high-energy cosmic ray (UHECR) spectrum and composition suggests a suppression of the flux below $\sim$1 EeV, as observed by the Pierre Auger Observatory and Telescope Array. A natural explanation for this phenomenon involves magnetic confinement effects. We investigate the possibility that UHECRs self-generate the magnetic turbulence necessary for such confinement via current-driven plasma instabilities.}
{Specifically, we show that the electric current produced by escaping UHECRs can excite a non-resonant streaming instability in the surrounding plasma. This instability reduces the diffusion coefficient in the source environment, effectively trapping particles with energies \(E \lesssim 0.6\) EeV \(\mathcal{L}_{45}^{1/2} R_{\text{Mpc}}^{-1} \lambda_{10}^{2}\) for times exceeding the age of the Universe. Here, \(\mathcal{L}_{45}\) is the source luminosity in units of \(10^{45}\) erg/s, \(R_{\text{Mpc}}\) is the radial size in Mpc, and \(\lambda_{10}\) is the intergalactic magnetic field coherence length in units of 10 Mpc. 
Here we discuss in detail the conditions, in terms of source luminosity, initial magnetic field, and the environment in which this complex phenomenon occurs, that need to be fulfilled in order for self-confinement to take place near a source of UHECRs, and the caveats that affect our conclusions. We emphasize that these conclusions are derived within a simplified model framework; their wider applicability requires that the assumptions hold in realistic source environments.}
{By modeling a population of UHECR sources with a luminosity function typical of extragalactic gamma-ray sources, we connect the spectrum of escaping particles to the luminosity distribution. Furthermore, we calculate the contribution of these confined particles to cosmogenic neutrino production, finding consistency with current observational constraints.
Our results suggest that self-induced turbulence may play an important role in shaping the UHECR spectrum and, in particular, may account for the flux suppression near their sources, offering a promising framework for interpreting current observations.}{}


\titlerunning{Excitation of the non-resonant streaming instability around
sources of Ultra-High Energy Cosmic Rays}
\authorrunning{A. Cermenati et al.}

\maketitle

\section{Introduction}
\nolinenumbers

The observed spectrum and composition of ultra-high-energy cosmic rays (UHECRs), as measured by the Pierre Auger Observatory (Auger)~\citep{abreu2021energy, PAO_AllPart_PhysRevD}, provide invaluable insights into their origin. The distribution of the depth of shower maximum ($X_{\rm max}$)~\citep{XmaxPRD2014, Hfrac2014PRD, Xmax2016PRB} reveals information about primary particle types, while spectrum fits, accounting for source evolution and propagation effects~\citep{Aab_2017JCAP, AbdulHalim_2023}, constrain key source parameters such as the spectral shape and elemental composition.

Several robust conclusions have emerged from these fits:
1) Above the second knee, UHECRs transition from a mix of protons and medium-mass nuclei (e.g., nitrogen) to a predominantly mixed composition near the ankle ($\sim 10^{18.3}$ eV), 2) At energies \( \gtrsim 10^{19} \) eV, the absence of a light component implies that the maximum rigidity of sources is limited to \( R \lesssim 10^{19} \) V, relaxing the demands in terms of acceleration models~\citep{Douglas_Bergman_2006, SHINOZAKI200418} ; 3) The inferred source spectrum is unusually hard, with slopes \( \gamma \lesssim 1 \), a result inconsistent with the \( \gamma \gtrsim 2 \) expected from standard acceleration models such as diffusive shock acceleration at non-relativistic and relativistic shocks~\citep{Blandford1978ApJ, BLANDFORD19871, Sironi2015}. Although unusual, such hard spectra have been found in some acceleration models, such as rapidly spinning neutron stars~\citep{Blasi2000, Arons2003, Kotera2015}, second-order stochastic acceleration in gamma-ray bursts~\citep{Asano2016}, and reconnection events in relativistic plasmas~\citep{Sironi2023}. More recent Auger data analyses even suggest inverted source spectra (\(\gamma < 0 \)) above the ankle, introducing a tension with existing astrophysical models~\citep{AbdulHalim_2023}. Furthermore, to reproduce observables below the ankle, an additional soft-spectrum component (\( \gamma \sim 3.5 \)) is needed, consistent with measurements of the proton spectrum by KASCADE-Grande~\citep{Kaskade2023ICRC} and IceTop~\citep{ICECUBE2019ICRC}.

A natural interpretation of these findings invokes confinement of UHECRs with energies below the ankle, a scenario that can be realized in two main classes of models: 1) Confinement around the sources: In this scenario, heavy nuclei experience energy-dependent confinement in the source vicinity or even inside the source. Their prolonged residence time enhances photo-disintegration processes, altering the observed composition and spectrum~\citep{UngerPRD, Muzio2022PRD}; 2) Magnetic Horizon Effect: Here, the delayed arrival of low-energy particles is attributed to diffusion in strong extragalactic magnetic fields (\( \gtrsim 10 \) nG). This results in an effective low-energy cutoff, dependent on the strength and structure of the intergalactic magnetic field~\citep{Aloisio_2004, MollerachJCAP, halim2024impact}.
Both models require specific assumptions about the source environment or the intergalactic magnetic field (IGMF). In particular, the latter demands field strengths near observational upper limits imposed by radio synchrotron and Faraday rotation measurements~\citep{Amaral2021, Carretti2022}.

A third possibility, originally proposed in~\citet{Blasi2015prl}, involves the generation of self-induced turbulence. The electric current carried by escaping UHECRs drives a non-resonant streaming instability~\citep{Bell2004mnras, BlasiAmatoKinetic}, which amplifies small-scale magnetic perturbations. These perturbations grow non-linearly, saturating at scales comparable to the gyroradius of the dominant particles in the current. This mechanism modifies UHECR transport near the source, effectively confining particles and regulating their escape. Notably, \citet{Blasi2015prl} demonstrated that for a source luminosity of \( \mathcal L \sim 10^{44} \)  erg/s, this self-confinement mechanism can trap sub-EeV particles for timescales exceeding the typical source lifetime of \( \sim 10 \)~Gyr. 

In this work, we build upon this last model and explore four critical aspects:
1) we investigate the minimum luminosity required for a UHECR source to induce a low-energy cutoff via self-confinement in the EeV energy range, as a function of the source parameters. Here, the term \emph{source} refers to the astrophysical object hosting the accelerator rather than the accelerator itself. For example, if gamma-ray bursts serve as the acceleration sites within a galaxy, the source in our context would be the galaxy. This distinction emphasizes the fact that it is the transport of UHECRs within the environment surrounding the astrophysical host, rather than around the acceleration region, that governs self-confinement.
2) We consider the advection of UHECRs with the background plasma, set in motion due to the accumulation of cosmic rays, following the approach outlined in~\citet{blasi2019prl}. Specifically, we find that the strong coupling between cosmic rays and the background plasma, driven by self-generated magnetic perturbations, imposes an upper limit on the current of particles escaping the source. Consequently, at sufficiently high source luminosities, the drift velocity of the plasma, assumed to be of the same order of magnitude as the Alfvén speed in the amplified magnetic field, becomes large enough to carry particles outside the confinement region on a timescale shorter than the source lifetime.
3) We review the constraints on the pre-existing magnetic field required for the growth of the non-resonant streaming instability~\citep{zweibel2010aas}. If the initial magnetic field is too strong, the non-resonant branch of the instability is suppressed. Conversely, if the field is too weak, the perturbations are induced on a scale comparable with the Larmor radius of background plasma protons, limiting the growth of the instability. The latter condition depends on the energy of the particles driving the current and generating the perturbations. However, this limitation is not severe if additional types of magnetic turbulence within the source suppress the escape of particles with energies \( \lesssim \)~PeV into the IGM. As previously noted, the source in this context refers to the astrophysical object hosting the accelerator, such as a galaxy or galaxy cluster, and thus the escape of low-energy particles can be regulated by turbulence not considered in this study.
4) As in these models, the source spectrum exhibits a non-linear dependence on source luminosity. We then model the escape of UHECRs from a population of putative sources with a specified luminosity function~\citep{Ajello_2009, Burlon_2011, Ajello_2012, qu2019} to investigate how this distribution impacts the observed spectrum at Earth. Notably, when self-generation effects are active, the spectrum of cosmic rays escaping the sources becomes nearly independent of the spectral shape at the acceleration site. Instead, the slope of the spectrum below the break depends on the shape of the source luminosity function.

Finally, we estimate the neutrino flux from confined protons near the source and compare it to the flux from escaping protons. Even under optimistic assumptions, we find that the confinement-dominated regions produce signals consistent with current observational limits~\citep{Kopper2017IceCubeLims, IceCube:2025ezc}.

While this work focuses on protons to highlight the self-confinement mechanism, the results are rigidity-dependent and extendable to mixed compositions, as required by UHECR observations~\citep{AbdulHalim_2023}. A detailed analysis of the implications of this model for the mass composition will be carried out in a forthcoming publication.

\section{Model}

\subsection{Excitation of the Non-Resonant Streaming Instability Around UHECR Sources}
\label{sec:instability}

The physical setup adopted here is similar to the one first introduced in Refs.~\citet{Blasi2015prl,blasi2019prl}: the source is assumed to have a typical radius $R$ and be located in a region in which a pre-existing turbulent magnetic field $B_0$ is present, with a typical coherence scale $\lambda_B$. UHECRs escaping the source gyrate around the direction of the local magnetic field and can be considered magnetized or unmagnetized depending on whether their gyroradius is smaller or larger than $\lambda_B$, respectively. For magnetized particles, we can safely assume that the guiding center follows the magnetic field line for at least a distance $\sim \lambda_B$ from the source, since under rather generic conditions on such scales, perpendicular diffusion is less relevant. 
We also assume that the differential spectrum of particles released into the surrounding medium is $\propto E^{-2}$ between a minimum ($E_{min}$) and maximum ($E_{max}$) energy. 
In this simplified picture, the number density of particles with energy larger than $E$ in the flux tube of length $\lambda_B$ can be written as
\begin{equation}\label{eq:nCR}
n_{\rm CR} (>E) = \frac{2 \mathcal L_{\rm p} E^{-1}}{c \Lambda \pi R_c^2 (E)},
\end{equation}
where \( \mathcal L_{\rm p} \) is the \emph{proton} luminosity of the source, \( \Lambda = \ln \left(\frac{E_{\rm max}}{E_{\rm min}}\right) \approx 20 \) serves as a normalization factor, and $R_c(E) =  R_L(E) + R$, with $R_L(E)$ being the Larmor radius of particles in the pre-existing magnetic field. Clearly, this expression relies on the assumption that the motion of the particles is approximately ballistic, with velocity $\sim c/2$, in the flux tube, as it is supposed to be in the absence of an appreciable level of magnetic perturbations. The introduction of $R_c(E)$ mimics the fact that the particles injected by the source are distributed on a transverse surface that is the largest between the size of the source, $R$, and the Larmor radius of the same particles in the pre-existing IGMF $B_0$. For reference values $B_0=1$ nG and $\lambda_{\rm B}=10$ Mpc, the transition from magnetized and unmagnetized particles occurs at $\sim$ 10 EeV. We assume $R\sim 1$ Mpc as a fiducial value of the source radius, but it is useful to keep in mind that this size may be appreciably smaller. Under these conditions, the size of the flux tube turns out to be roughly energy independent for energies $\lesssim 1$ EeV. It is worth noting that for these reference values of the parameters, an interesting phenomenology is expected to appear at the same energy where, based on the Auger data, one would expect the maximum energy of protons, mainly constrained by the mass composition at higher energies~\citep{AbdulHalim_2023}.

UHECRs escaping the source create a positive electric current density \( J_{\rm CR}(>E) = e \frac{c}{2} n_{\rm CR}(>E) \), which may excite a non-resonant streaming instability provided the energy density carried by the current is larger than the magnetic energy density in the pre-existing field~\citep{Bell2004mnras}:
\begin{equation}
\frac{E J_{\rm CR}(>E)}{e c} \gtrsim \frac{B_0^2}{4 \pi} \to \frac{\mathcal L_{\rm p}}{c \, \Lambda \pi R_c^2 (E)} \gtrsim \frac{B_0^2}{4 \pi} .
\label{eq:criticalcurrent}
\end{equation}

In the linear stage of the growth of the non-resonant instability, the modes that grow are quasi-purely growing at large wavenumbers. The maximum growth occurs at 
\begin{equation}
k_{\rm max} = \frac{4 \pi}{c} \frac{J_{\rm CR}(>E)}{B_0} \gg \frac{1}{R_L(E)},
\end{equation}
and the corresponding growth rate is 
\begin{equation}
\gamma_{\rm max}(E) = v_A k_{\rm max} \simeq \frac{2}{\sqrt{\pi n_b m_p}} \frac{e \mathcal L_{\rm p}}{ c \Lambda R^2 E},
\label{eq:growth_rate}
\end{equation}
where \( v_A = B_0 / \sqrt{4 \pi n_b m_p} \) represents the Alfvén speed in the surrounding magnetic field, and $n_b$ is the number density of the medium through which the cosmic rays propagate, which we assume to be of the order of the mean cosmological baryon density. Notice that, based on Equation~\eqref{eq:growth_rate}, the maximum growth rate is independent of the initial value of the local magnetic field, $B_0$.

In order for the instability to be excited, as discussed above, the condition in Eq.~\eqref{eq:criticalcurrent} must be fulfilled. 
This condition translates into an upper limit on the strength of the IGMF:
\begin{equation}
B_0 \lesssim B_\text{upper} = \sqrt{\frac{4 \mathcal L_{\rm p}}{\Lambda R^2 c}} \approx 25~\text{nG}
\left(\frac{\mathcal L_\text{p}}{10^{45} \, \text{ erg/s}}\right)^{1/2} \left(\frac{R}{\text{Mpc}}\right)^{-1}.
\label{eq:Bupper}
\end{equation}

Since the perturbations grow on small scales compared to the gyroradius of the particles dominating the current, particle transport is only weakly perturbed during this stage, and in fact, the magnetic field perturbations $\delta B$ on a scale $\sim k_{max}^{-1}$ can grow to values $\delta B/B_0 \gg 1$. At later stages, the $J\times B$ force, perpendicular to $B_0$, stretches the perturbations on larger scales. Eventually, the current becomes affected by the growth of the perturbations and the instability saturates, when $\delta B = \delta B_{\rm sat}\sim B_{\rm upper}$. Numerical simulations suggest that the time required for this saturation to be achieved requires $\sim 5-10$ e-folds~\citep{gargate2010aas}, which translates into a saturation time:
\begin{equation}
\tau_\text{sat}(E) \sim 5 \, \gamma_\text{max}^{-1}(E) \approx 2~\text{Gyr}~\left( \frac{\mathcal L_{\rm p}}{10^{45} \, \text{erg/s}} \right)^{-1} \left( \frac{R}{\rm Mpc} \right)^{2} \left( \frac{E}{\rm EeV} \right).
\label{eq:saturationtime}
\end{equation}

One should appreciate that, for UHECR with $E \lesssim $~EeV, the self-generated magnetic field reaches its saturation level $\delta B_{\rm sat}$ in a time much smaller than the duration of the source activity, $t_{\rm age} \sim 10$~Gyr. We stress once more that by source here we mean the astrophysical object from which UHECRs are escaping. Hence, the source duration can be reasonably assumed as comparable to the age of the universe, and $\mathcal L_p$ as the effective average luminosity. 

As first discussed in~\citet{zweibel2010aas}, the growth of the non-resonant instability depends on small-scale physics that is often overlooked. For our purposes, since $k_{max}R_L\gg 1$, it may happen that $k_{max}^{-1}$ becomes comparable with the Larmor radius of thermal ions, $R_{\rm L,th}$. The condition \( k_\text{max} R_\text{L,th} \lesssim 1 \) translates into a lower limit on the pre-existing IGMF:
\begin{equation}\begin{split}
B_0& > B_\text{lower} = \left( \frac{16 \mathcal L_{\rm p}^2 m_p k_B T_{\rm IGM}}{\Lambda^2 E_{\rm min}^2 R^4} \right)^{1/4} \\ \approx& 10^{-4}~\text{nG} \left( \frac{\mathcal L_{\rm p}}{10^{45} \, \text{erg/s}} \right)^{1/2} \left( \frac{T_{\rm IGM}}{10^4 \, \text{K}} \right)^{1/4} \left( \frac{R}{\rm Mpc} \right)^{-1} \left( \frac{E_{\rm min}}{\rm PeV} \right)^{-1/2}.
\end{split}
\label{eq:Blower}
\end{equation}

This bound is somewhat dependent upon the minimum energy of the particles that manage to reach the location where the growth rate is being estimated. The reference value shown in Eq.~\eqref{eq:Blower} is obtained by assuming that particles with energy $\gtrsim 1$ PeV reach a given location. 
It is highly non-trivial to assess this point: low-energy particles, well below the PeV range, may be potentially scattered by any type of small-scale turbulence, making it difficult to build a completely self-consistent picture. However, one could argue that it is highly unlikely for very low-energy particles to move quickly to large distances from the source, possibly due to other less effective instabilities (such as the resonant streaming instability, which is slower but still present). We will use the estimate in Eq.~\eqref{eq:Blower} as a warning to keep in mind when drawing conclusions concerning our model. 

In summary, if the IGMF lies within the range that supports the growth of the non-resonant instability, then within a timescale of order \( \tau_{\rm sat} \), the transport of UHECRs will be substantially altered by the emergence of a Bohm-like diffusion coefficient, which can be approximated as:
\begin{equation}
D = \frac{c}{3} \frac{E}{e \delta B_{\rm sat}} \approx 4~\frac{\text{Mpc}^2}{\text{Gyr}} \left(\frac{\mathcal L_\text{p}}{10^{45} \, \text{erg/s}}\right)^{-1/2} \left(\frac{R}{\text{Mpc}}\right) \left( \frac{E}{\rm EeV} \right).
\label{eq:Dbohm}
\end{equation}

As a consequence, for sufficiently low energies, the diffusion time out of the flux tube, characterized by a length scale \( \lambda_B \), can easily exceed the age of the universe. This behavior will be modeled in detail in the next section.

The fact that diffusion becomes very effective implies that cosmic rays are bound to the plasma and eventually follow the plasma motion. Moreover, the accumulation of cosmic rays near the source leads to large pressure gradients that set the background plasma in motion. This phenomenon was discussed in some detail in \citet{blasi2019prl}, where the authors provide an order of magnitude of the force exerted on the plasma and speculate that the plasma is set in motion with a velocity that is expected to be of the same order of magnitude as the Alfvén speed in the amplified field, where the field is argued to reach saturation primarily because of the Lorentz force reaction on the background plasma \citep{Bell2004mnras}, rather than due to the bulk motion of the plasma itself \citep{Riquelmeapj}.

Following the same line of thought as in \citet{blasi2019prl}, we assume that the plasma motion occurs with the Alfvén speed in the amplified field:
\begin{equation}
V_\text{A} \approx \frac{\delta B_\text{sat}}{\sqrt{4 \pi m_p n_b}} \sim 0.1~\frac{\text{Mpc}}{\text{Gyr}} 
 \left(\frac{\mathcal L_\text{p}}{10^{45} \text{ erg/s}}\right)^{1/2} \left(\frac{R}{\text{Mpc}}\right)^{-1},
\label{eq:alfvenspeed}
\end{equation}
and that cosmic rays are advected away from the source at about the same speed.

It is worth emphasizing that the displacement of the background plasma due to cosmic ray pressure gradients is not unexpected: it is in fact observed in hybrid-PIC simulations of the escape of particles from a source (e.g. a SNR) \citep{Schroer2022, Schroer2021} although in a different, less extreme, range of values of the relevant parameters. These simulations show that not only the plasma is set in motion in the direction of the local magnetic field, but also a cavity is excavated in such a plasma in the transverse direction. The plasma motion appears to occur at a speed comparable to the Alfvén speed in the amplified field.

\subsection{UHECR Transport in the IGM}

The non-linear chain of phenomena responsible for the excitation, growth, and saturation of the non-resonant instability, as well as the transport of cosmic rays in the self-generated perturbations, is extremely complex and poses a serious challenge not only from the computational point of view but from the conceptual point of view as well. For instance, after a time $\tau_\text{sat}(E)$, particles with energy $E$ are expected to be slowed down considerably, making their motion no longer ballistic. Consequently, one might argue that the previously proposed calculation of the current becomes inappropriate. However, as discussed in \citet{blasi2019prl}, while self-confinement makes transport diffusive, the current in the region where particles are confined remains unchanged at the lowest order: the current is conserved, but limited to a smaller region around the source, which is the reason why particles are confined and gradually leak outwards, provided there is sufficient time. 

Another complication arises from the fact that confinement causes the flux tube where UHECRs are confined to become over-pressured relative to the surrounding medium, so that the tube is expected to inflate a bubble, expanding at roughly the Alfv\'en speed \citep{Schroer2021, Schroer2022}. Given the relatively low value of such speed compared to the drift velocity of UHECRs due to diffusion, here we neglect this phenomenon as well, although it may become important in the lower energy region, where transport becomes dominated by advection. 

\begin{figure}[t]
    \centering
    \begin{subfigure}{0.48\textwidth}
    \includegraphics[width=\linewidth]{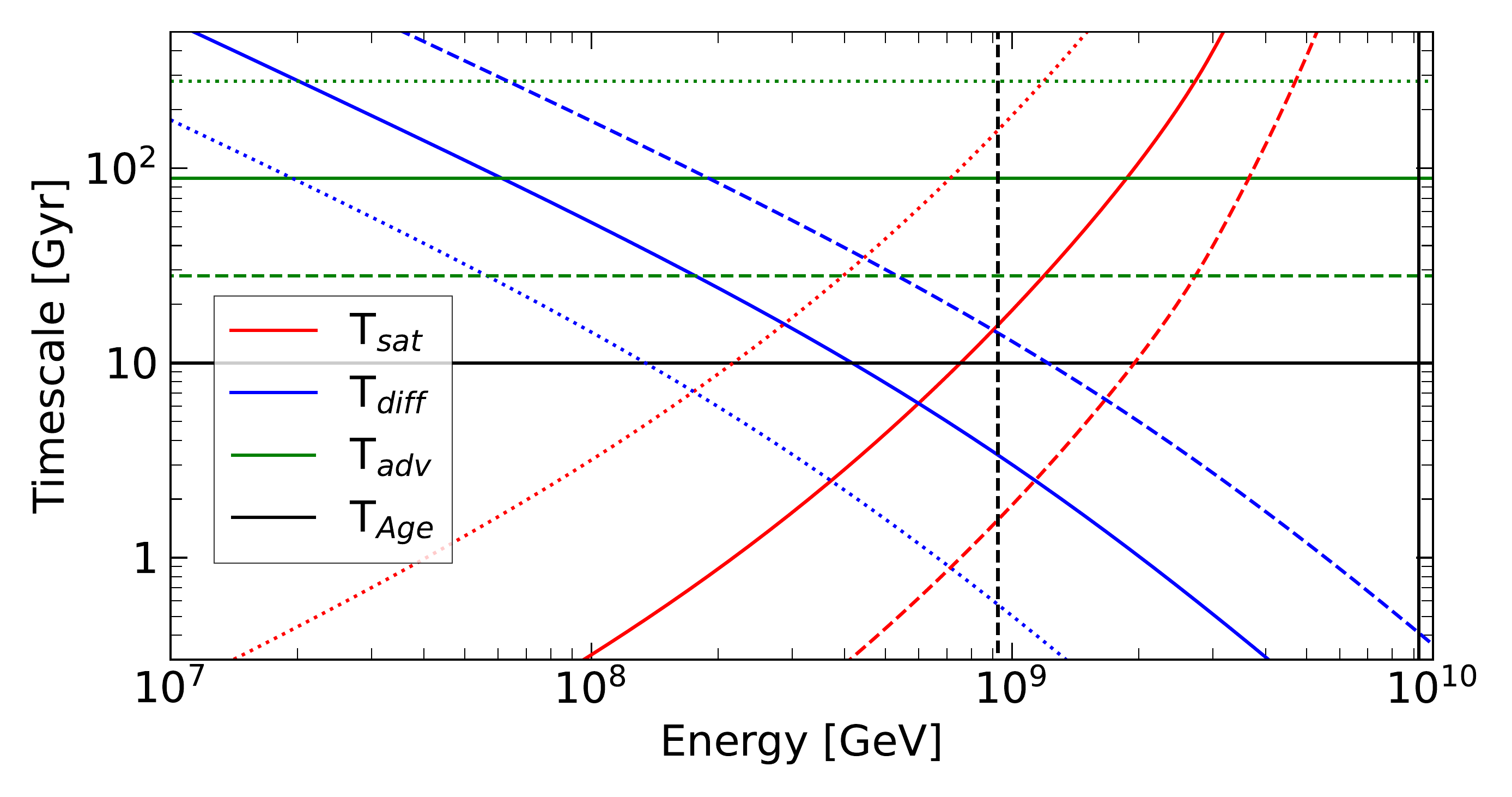}
    \caption{\label{fig:timescale}}
    \end{subfigure}
    \begin{subfigure}{0.48\textwidth}
    \includegraphics[width=\linewidth]{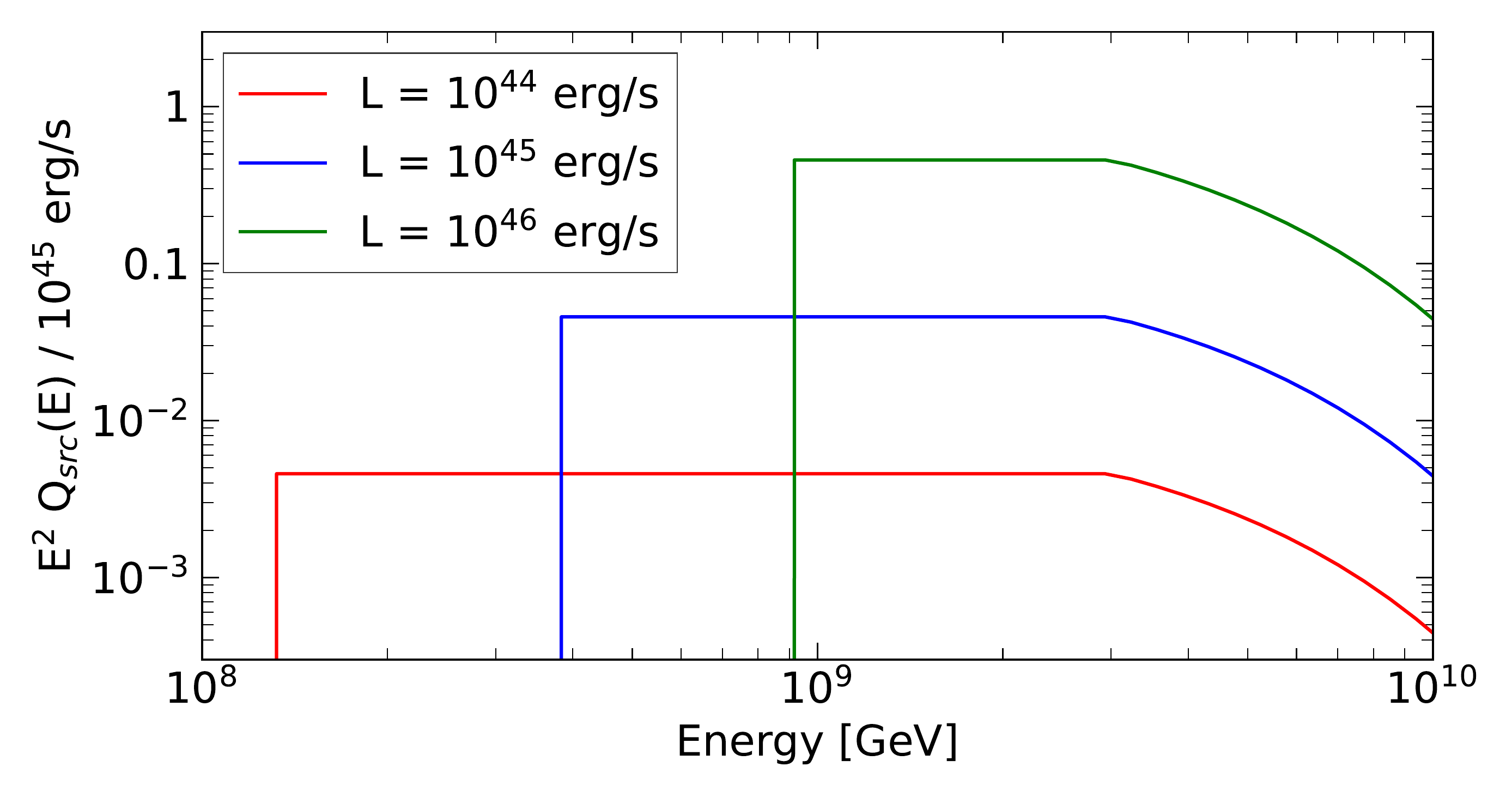}
    \caption{\label{fig:PSinjection}}
    \end{subfigure}
    \caption{Top panel: Relevant timescales for $\mathcal{L}_{\rm p}$ = $10^{44} \, {\rm erg/s} \, {\rm(dotted),} \, 10^{45} \, {\rm erg/s} \, {\rm (solid),} \, {\rm and} \, 10^{46} \, {\rm erg/s} \, {\rm (dashed)}$. The vertical lines show the critical energies defined by $R_L(E) = R$ (dashed) and $R_L(E) = \lambda_B$ (solid). Bottom panel: Cosmic rays injection (Eq.~\eqref{eq:Protons_PS_injection}) at t$_{\rm age}$ = 10 Gyr for $ \mathcal{L}_{\rm p} = 10^{44} \, {\rm erg/s} \, {\rm (red),} \, 10^{45} \, {\rm erg/s} \, {\rm (blue),} \, {\rm and} \, 10^{46} \, {\rm erg/s} \, {\rm (green)}$.\label{fig:figure1}}
\end{figure}

Given the paramount conceptual complexity of this scenario, here we treat the escape of UHECRs from the source vicinity in a very simple but physically motivated way: the spectrum of the escaping particles is the same as that injected by the source for those energies for which the escape time is shorter than $t_{\rm age}$ and vanishes otherwise.
We define the escape time from the near-source region as follows:
\begin{equation}
\tau_{\rm esc} (E) = \left( \frac{1}{\tau_{\rm adv}} + \frac{1}{\tau_{\rm diff}(E)} \right)^{-1},
\label{eq:escape_time}
\end{equation}
where $\tau_{\rm adv} \sim \frac{\lambda_B}{V_A}$ and $\tau_{\rm diff}(E) \sim \frac{\lambda_B^2}{4D(E)}$ are the advective and diffusive timescale, respectively. These time scales are shown in Fig.~\ref{fig:timescale} for $\lambda_B=10$ Mpc, $t_{\rm age}=10$ Gyr, and different values of the source luminosity.

Although the diffusion coefficient is expected to be Bohm-like (Eq. \eqref{eq:Dbohm}), at sufficiently high energy, the time required for the saturation of the non-resonant instability, $\tau_\text{sat}(E)$, will exceed $t_{\rm age}$. In this case, there will not be enough time for the generation of power on the resonant scale. Consequently, such particles will undergo small pitch angle scattering with a typical diffusion coefficient $D(E)\propto E^2$ \citep{Subedi2017}. To account for this effect, we write the diffusion coefficient in the following generalized form:
\begin{equation}
D = \frac{c}{3} \left(\frac{E_c}{e \delta B}\right) \left[ \left(\frac{E}{E_c}\right) + \left(\frac{E}{E_c}\right)^2 \right],
\end{equation}
where \(E_c\) is defined by the condition $\tau_\text{sat}(E)=t_{\rm age}$. Depending on specific parameters, $E_c$ is in the range $\approx$ (0.1-1) EeV. 

Moreover, when the particle energy becomes large enough that the gyration radius exceeds the size $R$ of the source, the effective cross-section of the flux tube increases as $E^2$, and it becomes much harder to excite the non-resonant instability. This occurs when $E \gtrsim E_R$, where $E_R \simeq eB_0R \simeq$ EeV $B_\text{nG} R_\text{Mpc}$. Finally, when particles cease to be magnetized on a scale $\lambda_B$, namely when $E \gtrsim E_M$ with $E_M = eB_0\lambda_B \simeq 10 \, \text{EeV} \, B_{\text{nG}} \lambda_{10}$, the escape of the particles becomes quasi-ballistic and the effect of self-generation does not take place. If the pre-existing magnetic field is on the order of $\sim 1$ nG, this is not too much of a concern since the range of energies where this occurs is close to or even in excess of the maximum energy of the protons compatible with the measured mass composition, namely a few EeV. However, for smaller strengths of the IGMF and if the coherence scale $\lambda_B$ is much smaller than 10 Mpc, this becomes a serious blow in that the propagation of particles in the range $\lesssim 1$ EeV becomes ballistic, and the low energy suppression disappears. Hence, to be on the safe side, the model discussed here requires the IGMF to be in the $\lesssim$ nG range, with $\lambda_B\gtrsim 10$ Mpc, for the non-resonant instability to be of phenomenological relevance in the EeV energy range. 

Provided these conditions are satisfied, the requirement that the instability is excited (Eq.~\eqref{eq:Bupper}) translates into a minimum source luminosity in the form of cosmic rays:
\begin{equation}
\mathcal{L}_{\rm p} \gtrsim \mathcal{L}_\text{min} \simeq 10^{42} \, \text{erg/s} \, \left( \frac{B_0}{1 \, \text{nG}} \right)^2 \, \left( \frac{R}{1 \, \text{Mpc}} \right)^2.
\label{eq:minimum_lum}
\end{equation}

It is worth noting that the Alfv\'en speed in the amplified field increases with the amplified magnetic field, which typically happens when the source luminosity is higher. This increase would reduce the advection time, which may eventually become too short in the EeV range to guarantee CR confinement. The condition that the advection time exceeds the source age, $V_A t_\text{age} \gtrsim \lambda_B$, sets an upper limit on the source luminosity for which the considerations discussed here apply:

\begin{equation}
\mathcal{L}_{\rm p} \lesssim L_\text{max} \simeq 10^{47} ~\frac{\text{erg}}{\text{s}} \left( \frac{\lambda_B}{\text{10 Mpc}} \right)^2 \left( \frac{R}{\text{1 Mpc}} \right)^2 .
\label{eq:maximum_lum}
\end{equation}

For typical values of the parameters, this upper limit does not pose a severe concern for the applicability of our calculations. 

When the source luminosity is within the range bounded by the values set by Equations~\eqref{eq:minimum_lum} and~\eqref{eq:maximum_lum}, cosmic ray confinement near the source occurs if the diffusion time exceeds the source age. This phenomenon results in a \emph{cutoff} in the spectrum of UHECRs released into the IGM at energies below

\begin{equation}
E_\text{D} \simeq 0.6 \, \text{EeV} \,
\left(\frac{\mathcal{L}_\text{p}}{10^{45} \, \text{erg/s}}\right)^{1/2} 
\left(\frac{R}{\text{Mpc}}\right)^{-1}
\left(\frac{\lambda_\text{B}}{10 \, \text{Mpc}}\right)^2 .
\label{eq:Ecut}
\end{equation}

Modeling the exact shape of this cutoff is beyond the scope of this work and it would not significantly alter the conclusions. Instead, we adopt a simple yet effective approach that captures the essential physics of the problem. In this context, we model the released spectrum of the source as:
\begin{equation}
Q_{\text{src}}(E,\mathcal L_{\rm p}) = q(E) \mathcal{H}\left[ t_{\rm age} - \tau_{\rm esc}(E,\mathcal L_{\rm p}) \right],
\label{eq:Protons_PS_injection}
\end{equation}
where \( q(E)= \frac{\mathcal L_\text{p} E^{-2}}{\Lambda} \), and the Heaviside function \( \mathcal{H} \) ensures that particles with escape times longer than the source age remain confined within the source. The shape of the spectrum of escaping particles from the near-source region is shown in Fig.~\ref{fig:PSinjection} for three values of the source luminosity, normalized to E$_\text{min}$ $=$ 1 GeV, and for our reference values of the environmental parameters, $R=1$ Mpc, $\lambda_B=10$ Mpc, $B_0=1$ nG and E$_\text{max}=$ 3 EeV.

\section{Results}
\label{sec:protons}

\subsection{Emissivity from a population of non-identical sources}
\label{sec::population}

As discussed in the previous section, the spectrum of a source that is actually released into the IGM depends in a non-linear way upon the source luminosity. Therefore, it is interesting to explore what happens when a luminosity function $\Phi (\mathcal{L}_{\rm p}, z)$ is assumed for the putative source population. The case of identical sources with a single value of the luminosity, usually adopted in the literature on the origin of UHECRs for simplicity (see for instance~\citealt{AbdulHalim_2023}), can be recovered by restricting the luminosity function to a narrow range. 

The emissivity in the form of protons as a function of energy at a given redshift can be easily calculated as:
\begin{equation}
Q_p(E,z) = \int_{\mathcal{L}_\text{low}}^{\mathcal{L}_\text{high}} d\mathcal{L}_{\rm p} \, \Phi(\mathcal{L}_{\rm p},z) \, Q_{\rm src}(E, \mathcal{L}_{\rm p}) \, ,
\label{eq:cosmological_emissivity}
\end{equation}
where $\mathcal{L}_\text{low}$ and $\mathcal{L}_\text{high}$ identify the minimum and maximum luminosity for which the luminosity function applies. 

The simplest choice of a luminosity function, useful to clarify some points, is a power law distribution:
\begin{equation}
    \Phi_\text{PL} (\mathcal{L}_{\rm p}, z) = \frac{dN}{d\mathcal{L}_{\rm p} dV} = A \times \begin{cases}
        & \left( \frac{\mathcal{L}_{\rm p}}{\mathcal{L_\text{low}}} \right)^{-\beta} \, \text{if} \, \mathcal{L}_\text{low} < \mathcal{L}_{\rm p} < \mathcal{L}_\text{high} \\
        & 0 \, \text{ otherwise,}
    \end{cases}
\end{equation}
where \( V \) is the comoving volume and $A$ is a normalization constant with units of Mpc$^{-3}$ erg$^{-1}$ s. 

\begin{figure}[t]
    \centering
    \begin{subfigure}{0.48\textwidth}
    \includegraphics[width=\linewidth]{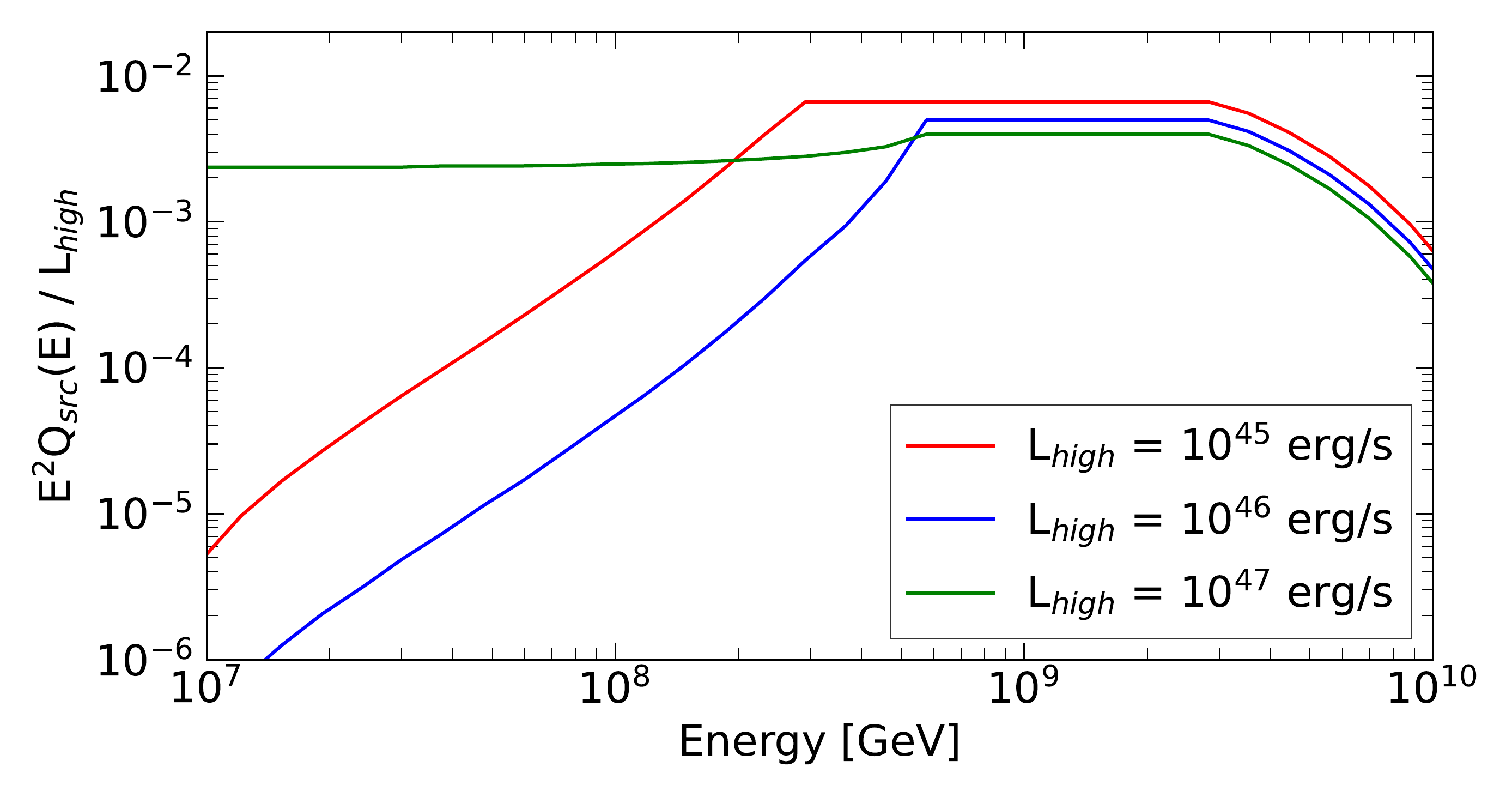}
    \caption{\label{fig:SPL_Lmax}}
    \end{subfigure}
    \hfill
    \begin{subfigure}{0.48\textwidth}
    \includegraphics[width=\linewidth]{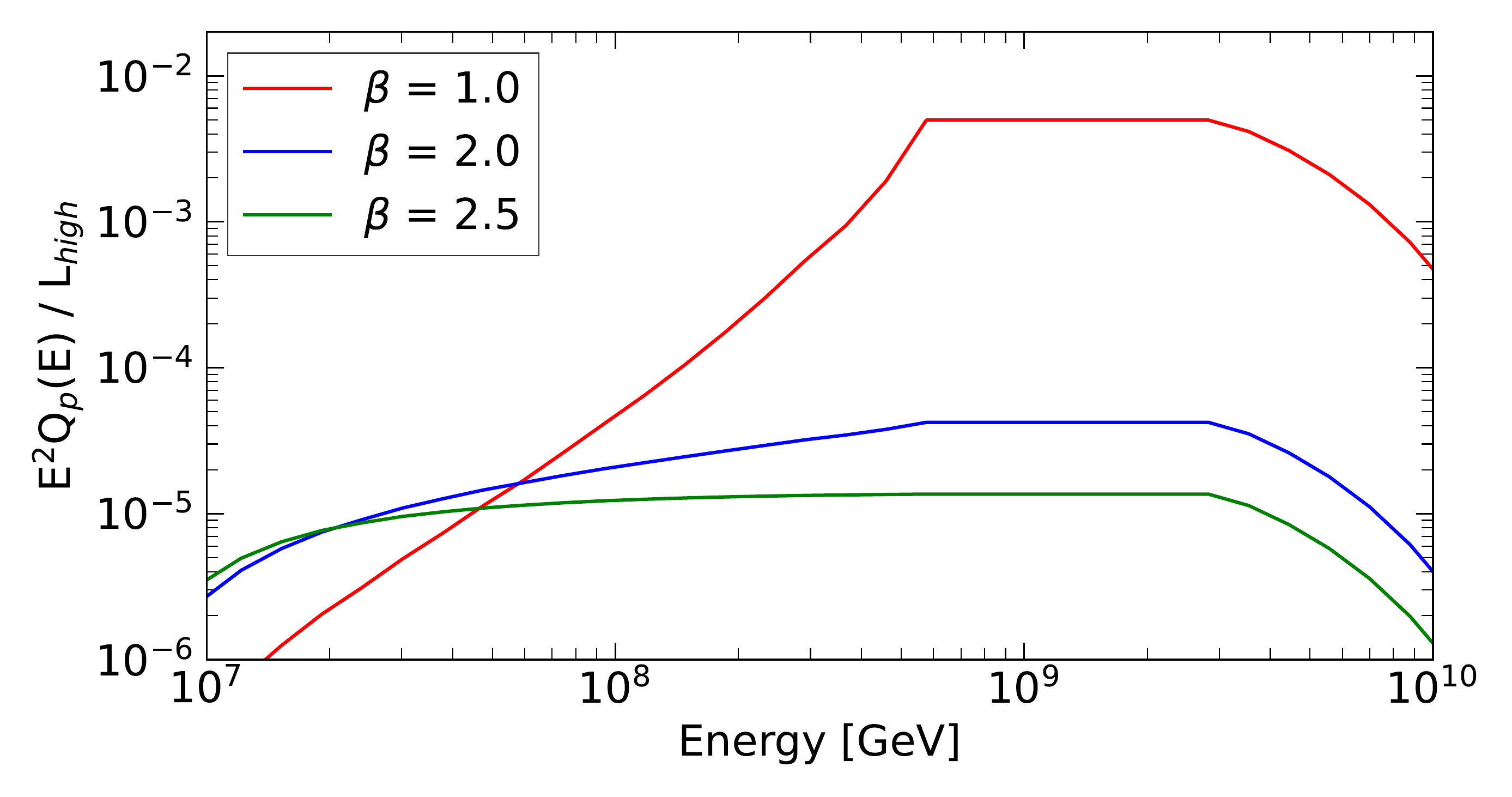}
    \caption{\label{fig:SPL_beta}}
    \end{subfigure}
    \hfill
    \begin{subfigure}{0.48\textwidth}
    \includegraphics[width=\linewidth]{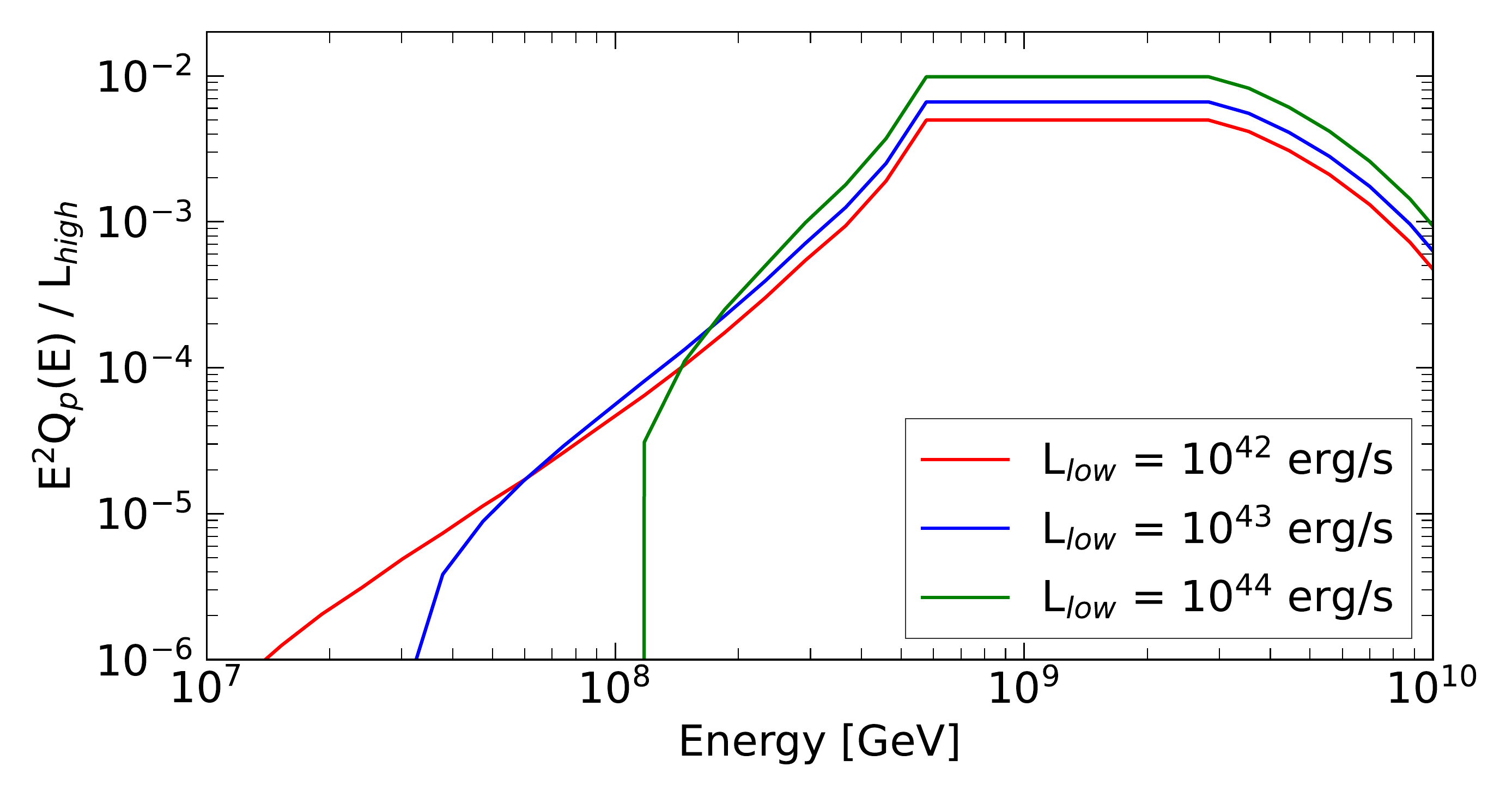}
    \caption{\label{fig:SPL_Lmin}}
    \end{subfigure}
    \caption{Proton emissivity calculated from Eq.~\eqref{eq:cosmological_emissivity}, considering a simple power law Luminosity Function. Top panel: different choices of $\mathcal{L}_{\rm high}$, with $\beta = 1$ and $\mathcal{L}_{\rm low} = 10^{42}$ erg/s. Central panel: different choices of $\beta$, with $\mathcal{L}_{\rm low} = 10^{42}$ erg/s and $\mathcal{L}_{\rm high} = 10^{46}$ erg/s. Bottom panel: different choices of $\mathcal{L}_{\rm low}$, with $\beta = 1$ and $\mathcal{L}_{\rm high} = 10^{46}$ erg/s.  \label{fig:figure2}}
\end{figure}

The emissivity in the form of protons, given a power law luminosity function, is shown in Figure~\ref{fig:figure2}. In the left panel, the results for $\beta=1$ are presented for three different values of $\mathcal{L}_\text{high}$, as indicated. Notice that for $\beta<2$, the emissivity at a given energy is dominated by the luminosity region near $\mathcal{L}_\text{high}$, which is the reason why in the left panel we only change the value of $\mathcal{L}_\text{high}$. The suppression of the flux of protons injected at low energies is easily identifiable and is due to the self-confinement of UHECR near the sources. The suppression is not as sharp as the one in Figure~\ref{fig:figure1} as a result of the convolution of different source luminosities. For $\beta<2$, it can be easily shown that the energy dependence of the emissivity for self-confined particles is approximately $\mathcal{Q}(E) \propto E^{-2 + 2(2-\beta)}$. This mimics a source spectrum that is harder than the canonical $E^{-2}$, although this shape does not directly relate to the acceleration process, as discussed earlier. The drop in the emissivity for energies $\gtrsim 3$ EeV is due to a cutoff in the source spectrum at that energy, required to fit the Auger data as found in previous phenomenological descriptions (see for instance~\citealt{Aloisio_2004,AbdulHalim_2023}). The curve corresponding to $\mathcal{L}_\text{high}=10^{47}$ erg/s shows a plateau at low energies (no suppression). This case highlights the role of advection: for high source luminosities, the CR current is so intense that the diffusion coefficient becomes very small, causing particles to escape due to advection with the self-generated perturbations. This case has been shown only as an illustration of the role of advection, but sources with such a high CR luminosity may imply a possibly unrealistic total luminosity of the sources, at least when averaged over a time $t_{\rm age}\sim 10$ Gyr. 

In the central panel of Figure~\ref{fig:figure2}, we present the emissivity for $\mathcal{L}_\text{low} = 10^{42}$ erg/s and $\mathcal{L}_\text{high} = 10^{46}$ erg/s, with three values of $\beta$. For $\beta=1$, there is a noticeable flux suppression for $E\lesssim 0.5$ EeV, as discussed previously. In the case of $\beta=2.5 (>2)$, the main contribution to the emissivity is provided by low luminosity sources, for which the self-confinement is limited to lower energy CRs, as illustrated by the green curve in the central panel of Figure~\ref{fig:figure2}. Given the small value of $\mathcal{L}_\text{low}$, the suppression occurs at exceedingly low energies, approximately 10 PeV. 

Finally, in the right panel of Figure~\ref{fig:figure2}, we show the emissivity in the form of protons for $\beta=1$ and different values of $\mathcal{L}_\text{low}$, while $\mathcal{L}_\text{high}$ is fixed to the value of $10^{46}$ erg/s. As in the previous cases, the high energy suppression is due to the adoption of an exponential cutoff at $3$ EeV. Higher values of $\mathcal{L}_\text{low}$ result in a more pronounced suppression of the emissivity at higher energies compared to lower $\mathcal{L}_\text{low}$ values. As a consequence, the region of hardening in the emissivity, due to the convolution with the luminosity function, extends to a narrower energy region when $\mathcal{L}_\text{low}$ increases. Nonetheless, in all scenarios, the effect of self-confinement, together with the convolution with a luminosity function, leads to mimicking a hard injection spectrum, again unrelated to the intrinsic spectrum of accelerated particles. 

The case of a simple power-law luminosity function, as illustrated above, provides useful insights into the expected effects: luminous sources cause self-confinement of UHECRs that extends to higher energies, and the convolution with a luminosity function, for $\beta<2$, leads to a hardening below the energy for which confinement is caused by sources with luminosity around $\mathcal{L}_\text{high}$. In this energy range, the spectrum behaves as $\mathcal{Q}(E) \propto E^{-2 + 2(2-\beta)}$. For $\beta>2$, the confinement is dominated by the sources with the lowest luminosity. 

\begin{figure}[t]
    \centering
    \begin{subfigure}{0.48\textwidth}
    \includegraphics[width=\linewidth]{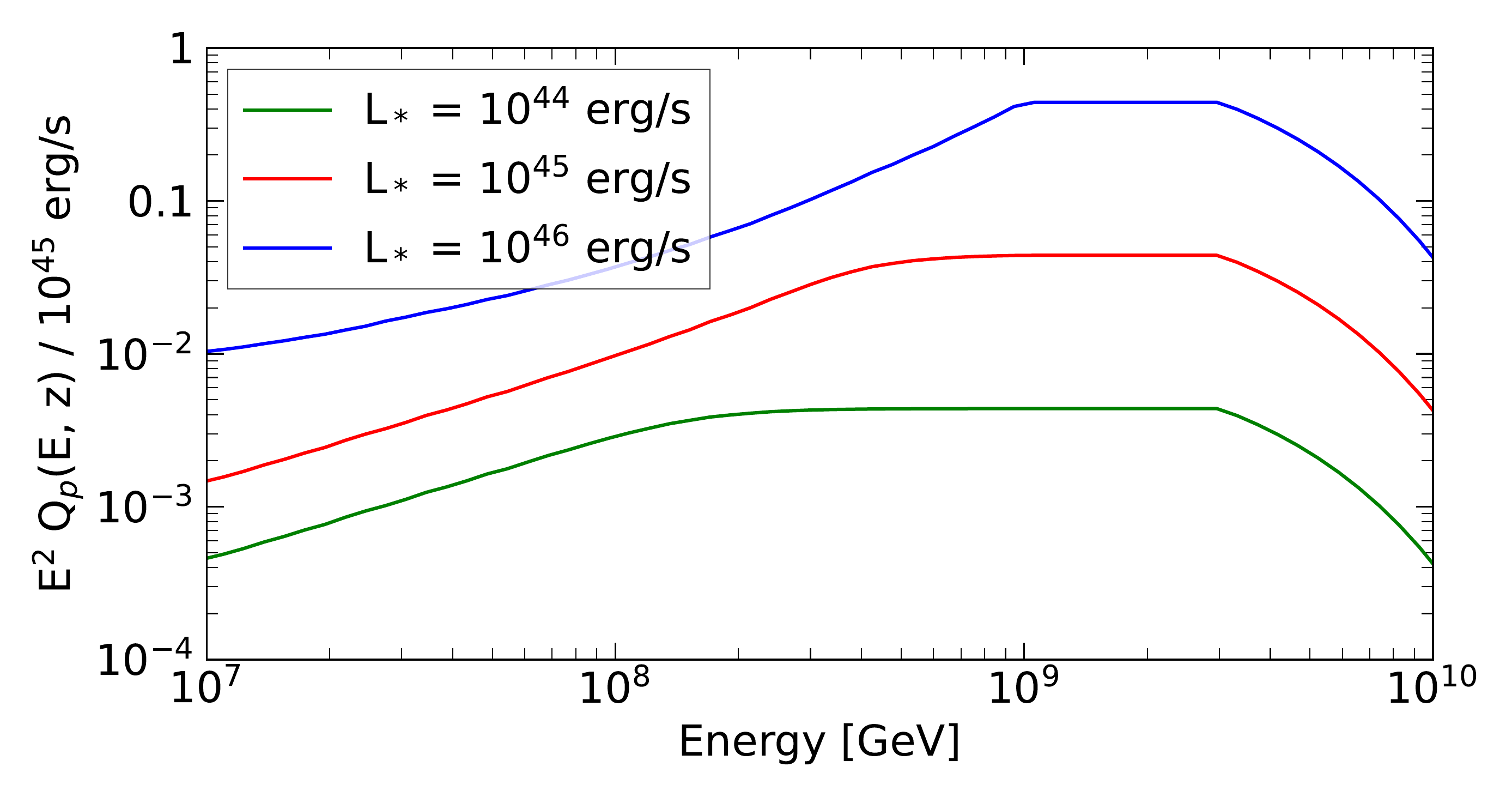}
    \caption{\label{fig:BPL_injection_lum}}
    \end{subfigure}
    \hfill
    \begin{subfigure}{0.48\textwidth}
    \includegraphics[width=\linewidth]{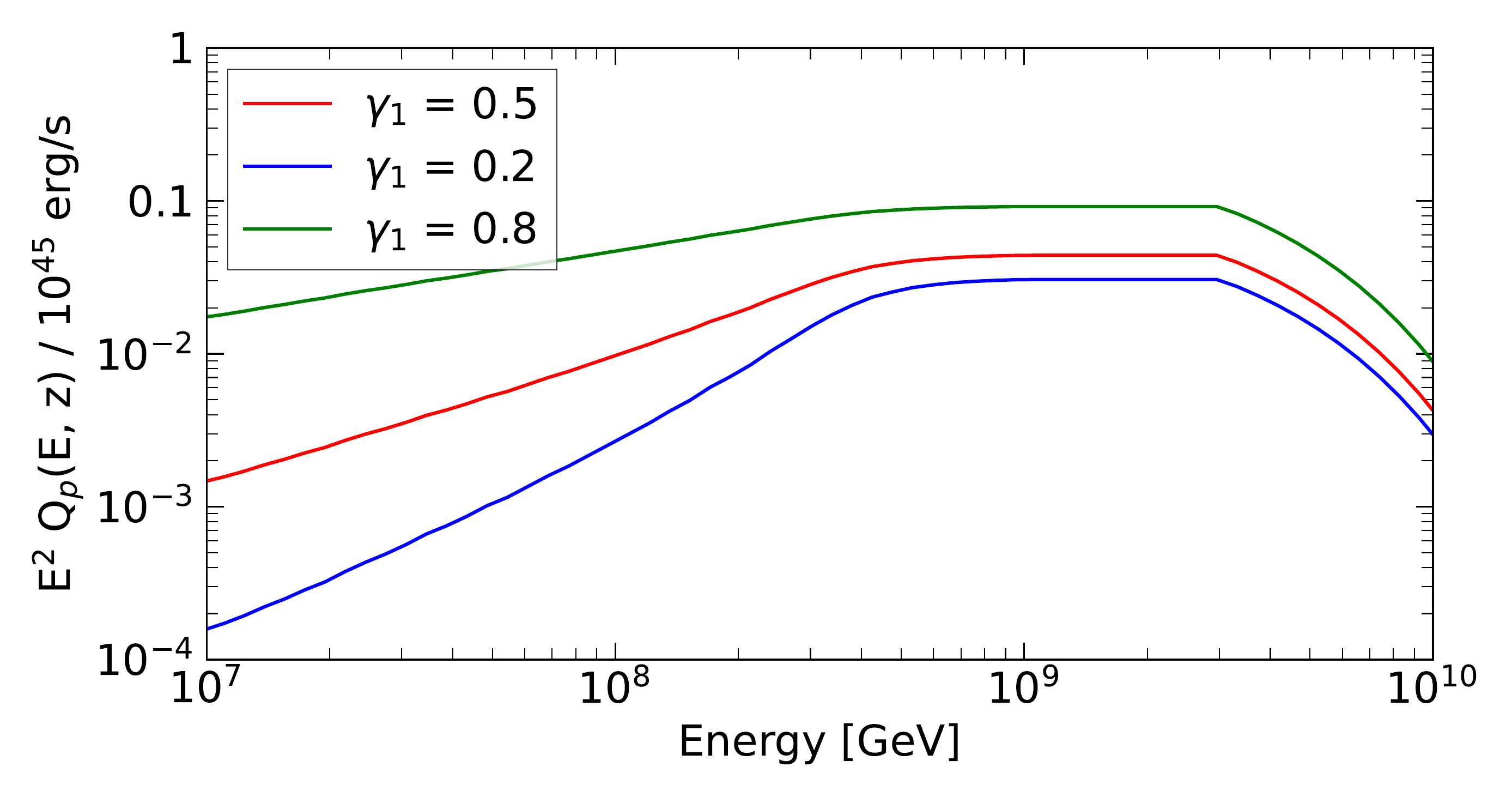}
    \caption{\label{fig:BPL_injection_slope}}
    \end{subfigure}
    \hfill
    \begin{subfigure}{0.48\textwidth}
    \includegraphics[width=\linewidth]{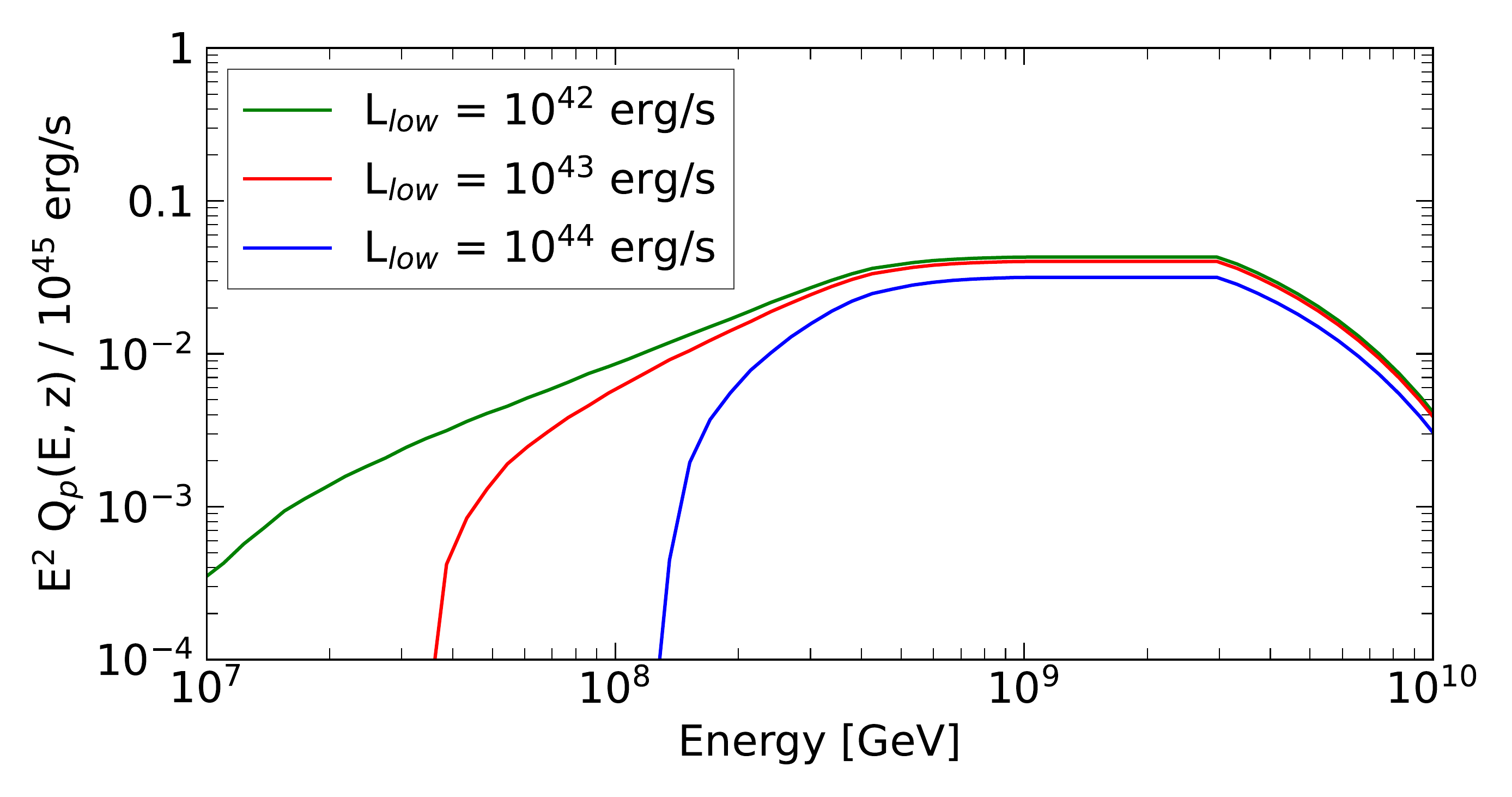}
    \caption{\label{fig:BPL_injection_lumMin}}
    \end{subfigure}
    \caption{Proton emissivity calculated from Eq.~\eqref{eq:cosmological_emissivity}, considering a broken power law Luminosity Function (Eq.~\eqref{eq:LF_shape}) with $\gamma_2 = 2.5$. Top panel: different positioning of $\mathcal{L}_*$; the low luminosity index is $\gamma_1 = 0.5$. Central panel: different low luminosity behavior; the positioning of the break is fixed at $\mathcal{L}_* = 10^{45}$ erg/s. Bottom panel: different choices of $\mathcal{L}_{\rm low}$; the position of the break is $\mathcal{L}_* = 10^{45}$ erg/s and the slope is $\gamma_1 = 0.5$. \label{fig:figure3}}
\end{figure}

Although this case may be sufficient to support the physical picture that we are interested in, since in many cases the luminosity function is modeled as a broken power law, for instance in the cases of AGN and Starburst Galaxies as sources or astrophysical objects hosting the sources, we briefly discuss this scenario using the following functional form for the luminosity function:
\begin{equation}
\Phi(\mathcal L_{\rm p},z) = \frac{dN}{d\mathcal L_{\rm p} dV} = \frac{A e(z)}{\log(10) \mathcal L_{\rm p}} \left[ \left( \frac{\mathcal L_{\rm p}}{\mathcal L_*} \right)^{\gamma_1} + \left( \frac{\mathcal L_{\rm p}}{\mathcal L_*} \right)^{\gamma_2} \right]^{-1},
\label{eq:LF_shape}
\end{equation}
where \( V \) is the comoving volume, \( \gamma_i \) are the slopes characterizing the low and high-luminosity behavior, \( L_* \) is a break luminosity, and \( A \) is a normalization constant. The term \( e(z) \) can be chosen to model the redshift evolution of the luminosity function, which varies with the source population. Here, for simplicity and for the sole purpose of illustrating the physical results, we adopt a constant redshift evolution $e(z) = 1$ up to \( z_{\rm max} = 3 \). A more detailed review of the luminosity distribution modeling for galaxies and AGN-like objects can be found in~\citet{Fotopoulou2016}.
As in the previous case, the emissivity in the form of protons can be computed using Equation~\eqref{eq:cosmological_emissivity}. 
We show our results in Figure~\ref{fig:figure3} for different choices on the parameters defining the luminosity function.

Given the wide variety of possible choices of the values of the parameters, we try to find guidance in the luminosity function for some sources, such as that of AGN-like sources, for which typically values of $\gamma_1 \lesssim 1$ and $\gamma_2 \gtrsim 2$ are fitted in the X-ray~\citep{Ajello_2009, Burlon_2011} and $\gamma$-ray~\citep{Ajello_2012, qu2019} bands. 
Based on the results found in the cases of a power law luminosity function, it is clear that most results of physical relevance will be determined by the choice of $\gamma_1$ and $\mathcal{L}_*$. Therefore, we fix $\gamma_2=2.5$ and briefly comment on the dependence of the results on the other parameters. Again, we can use our findings in the power law case as guidance: most of our previous conclusions still hold when the role of $\mathcal{L}_\text{high}$ is now played by $\mathcal{L}_*$ and the role of $\beta$ is played by $\gamma_1<1$. In this way, the interpretation of the three panels in Figure~\ref{fig:figure3} is straightforward. As in the previous case, the high-energy exponential drop in the emissivity is due to the assumed maximum energy of protons in the accelerator. 

\subsection{Diffuse proton flux}

Here we focus on the spectrum of protons at the Earth, when the source spectrum is computed accounting for self-confinement, as discussed above. The propagated flux of protons at \( z=0 \) can be calculated following~\citet{Berezinsky_Gazizov_prd} and it reads:
\begin{equation}
J_p (E) = \frac{c}{4\pi} \int_{0}^{z_{\text{max}}} dz \left| \frac{dt}{dz} \right| Q_p(E_g(E,z),z) \frac{dE_g(E, z)}{dE},
\end{equation}
where \( \left| \frac{dt}{dz} \right| = H^{-1}(z)/(1+z) \), and the Jacobian \( \frac{dE_g}{dE} \) is determined by the energy losses suffered by the protons during transport. We include pair production and photo-pion production on the Cosmic Microwave Background (CMB) radiation field as channels of energy losses and we compute the corresponding energy loss rates as functions of both energy and redshift by interpolating numerical tables generated using the latest release of the SimProp code~\citep{SimProp2017}. The adopted cosmology is the Planck $\Lambda$CDM model, with $h = 0.67$, $\Omega_m = 0.32$, and $\Omega_\Lambda = 0.68$~\citep{Planck2015}. 

\subsection{Diffuse Neutrinos from UHECR Sources}

In the standard scenario, UHECRs are promptly injected by sources into the IGM and propagate over cosmological distances. During their journey, interactions with background radiation fields generate a cosmogenic neutrino flux, which is directly tied to the flux of protons reaching the Earth.

In contrast, the scenario discussed in this work (like other scenarios involving near-source confinement~\citep{UngerPRD, Muzio2022PRD}) introduces an additional contribution to the diffuse neutrino background. This contribution arises from UHECRs that are unable to escape their near-source regions. In our framework, protons confined within the source environment cannot propagate into the IGM but may still produce neutrinos via photopion production.

Due to the relatively low maximum energies allowed by current mass composition measurements, most of these interactions occur with the Extragalactic Background Light (EBL) rather than with the CMB alone. 
For the EBL, we adopt the most recent determination provided by~\citet{saldana2021ebl}. While the local EBL near sources could, in principle, exceed the average EBL, we neglect this effect here, as it strongly depends on the specific type of source under consideration.

Below, we describe the calculation of the neutrino flux produced by confined UHECRs, \( J_\nu^{\rm conf} \), and compare it to the purely cosmogenic neutrino flux, \( J_\nu^{\rm cosmo} \), which arises from UHECRs that successfully escape the near-source regions.

In the calculations of the effect of photopion production, we made use of the parametrization of the differential cross section described in~\citet{Kelner2008prd}, which is obtained by fitting the energy distribution of the photopion interactions resulting from particle physics simulations performed with SOPHIA~\citep{Sophia_code}. The energy distribution of the outgoing neutrinos is parametrized in terms of the fraction of energy of the impinging proton taken by the emitted neutrino, $x = E_\nu/E_p$. Using the same formalism adopted in~\citet{Kelner2008prd}, the number of neutrinos produced per unit time and energy in the entire volume defined by the flux tube around a source is given by:
\begin{equation}
    q_\nu^\text{conf}(E_\nu,z) = \int_{E_\nu} \frac{dE_p}{E_p} N_p(E_p, t_\text{age}) \mathcal{R}(E_p, E_\nu, z),
\end{equation}
where $N_p(E,t)$ is the number of protons with energy $E$ confined within the flux tube at the time $t$, and $\mathcal{R}(E_p, E_\nu, z)$ is the production rate of neutrinos with energy $E_\nu$ from protons with energy $E_p$ interacting with the background photons at redshift $z$. The production rate is defined using the parametrized cross-section as:
\begin{equation}
\mathcal{R}(E_p, E_\nu, z) = c \int_{\epsilon_\text{th}(E_p)} \! d\epsilon \, n_\gamma(\epsilon, z) \,  \frac{d\sigma}{dx}(E_p, E_\nu, \epsilon),
\end{equation}
where $\epsilon_\text{th}(E_p)$ identifies the threshold energy for the photopion interaction. To estimate the number of protons confined within the flux tube, we consider a continuous injection of protons from the source at a rate $q(E)$ and continuous escape from the flux tube at a rate defined by Equation~\eqref{eq:escape_time}:
\begin{equation}
\frac{\partial N_p(E,t)}{\partial t} = - \frac{N_p(E,t)}{\tau_\text{esc}(E)} + q(E).
\end{equation}

The solution at $t = t_\text{age}$ holds:
\begin{equation}
N_p(E_p,t_\text{age}) = q(E_p) \tau_\text{esc}(E_p) \left[ 1 - e^{-\frac{t_\text{age}}{\tau_\text{esc}(E_p)}} \right],
\end{equation}
which depends non trivially on the source luminosity through the protons escape time. The cosmological neutrino emissivity from confined protons is obtained by summing over the contribution of all the sources:
\begin{equation}
Q_\nu^\text{conf}(E_\nu, z) = \int \! d\mathcal{L}_{\rm p} \, \Phi(\mathcal{L}_{\rm p}, z) \, q_\nu^\text{conf} (E_\nu, \mathcal{L}_{\rm p}, z),
\label{eq:Qnu}
\end{equation}
Finally, the neutrino flux at Earth is obtained by solving the cosmological transport equation:
\begin{equation}
J_\nu^\text{conf} (E_\nu) = \frac{c}{4\pi} \int_0^{z_\text{max}} dz \, \left| \frac{dt}{dz} \right| \, Q_\nu^\text{conf}(E_\nu (1 + z), z) \, (1 + z).
\label{eq:cosmo_nu_flux}
\end{equation}

On the other hand, the flux of \emph{cosmogenic} neutrinos is generated by unconfined protons propagating over cosmological distances. The proton spectral density at arbitrary redshift is determined in a way similar to~\citet{Berezinsky_Gazizov_prd}:
\begin{equation}
n_p(E,z) = \int_{z}^{z_{\text{max}}} dz_g \left| \frac{dt}{dz_g} \right| Q_p(E_g(E,z, z_g),z_g) \frac{dE_g(E, z, z_g)}{dE}.
\end{equation}
Hence, the cosmogenic neutrino emissivity reads:
\begin{equation}
    Q_\nu^\text{cosmo} (E_\nu,z) = \int_{E_\nu} \frac{dE_p}{E_p} n_p(E_p,z)\mathcal{R}(E_p,E_\nu,z),
\label{cosmo_nu_injection}
\end{equation}
and the diffuse cosmogenic neutrino flux at Earth is calculated as in Eq.~\eqref{eq:cosmo_nu_flux}:
\begin{equation}
J_\nu^\text{cosmo} (E_\nu) = \frac{c}{4\pi} \int_0^{z_\text{max}} dz \, \left| \frac{dt}{dz} \right| \, Q_\nu^\text{cosmo}(E_\nu (1 + z), z) \, (1 + z).
\end{equation}

\begin{figure}[t]
\centering
    \begin{subfigure}{0.48\textwidth}
    \includegraphics[width=\linewidth]{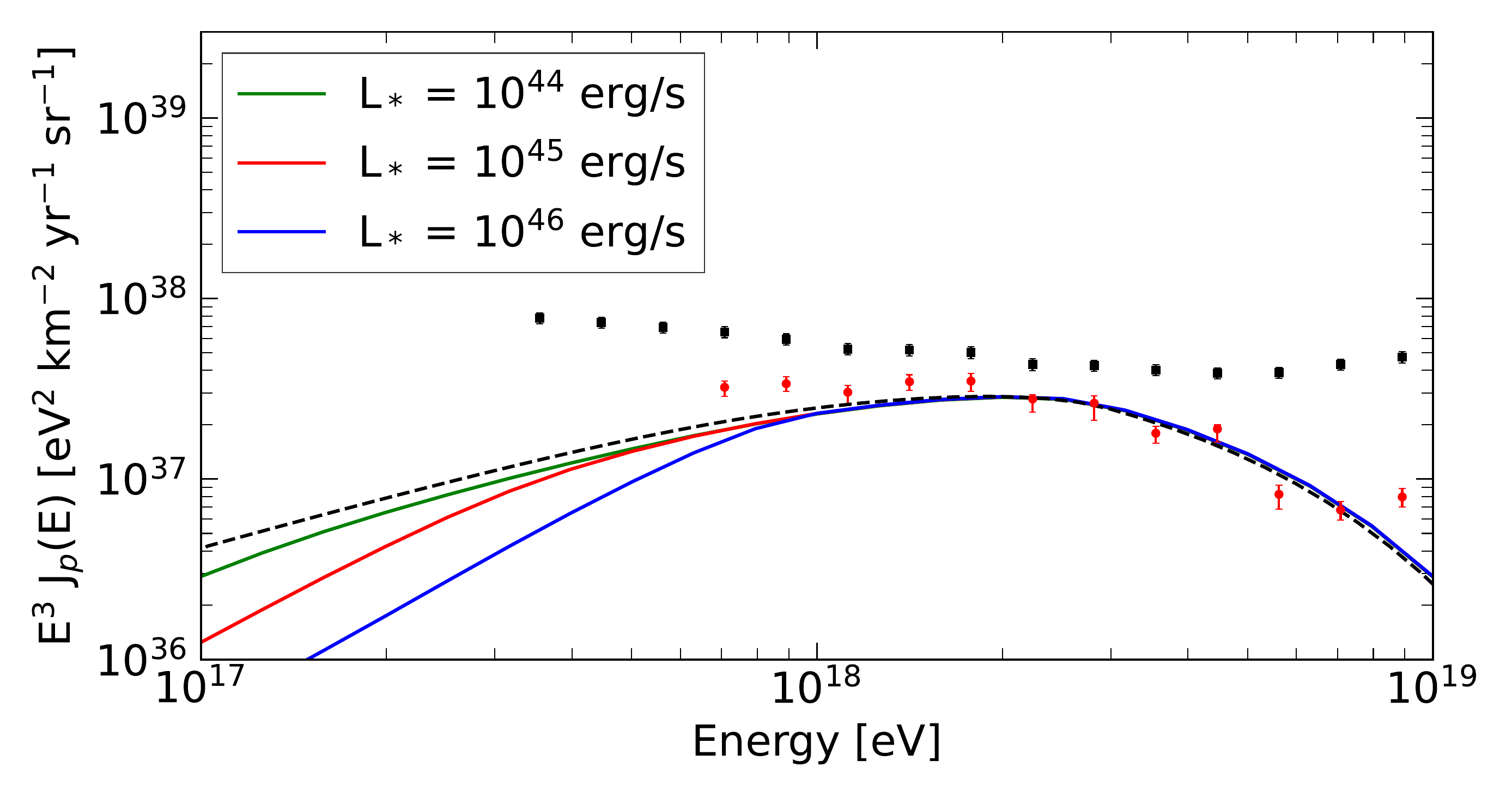}
    \caption{\label{fig:BPL_spectrum_lum}}
    \end{subfigure}
    \hfill
    \begin{subfigure}{0.48\textwidth}
    \includegraphics[width=\linewidth]{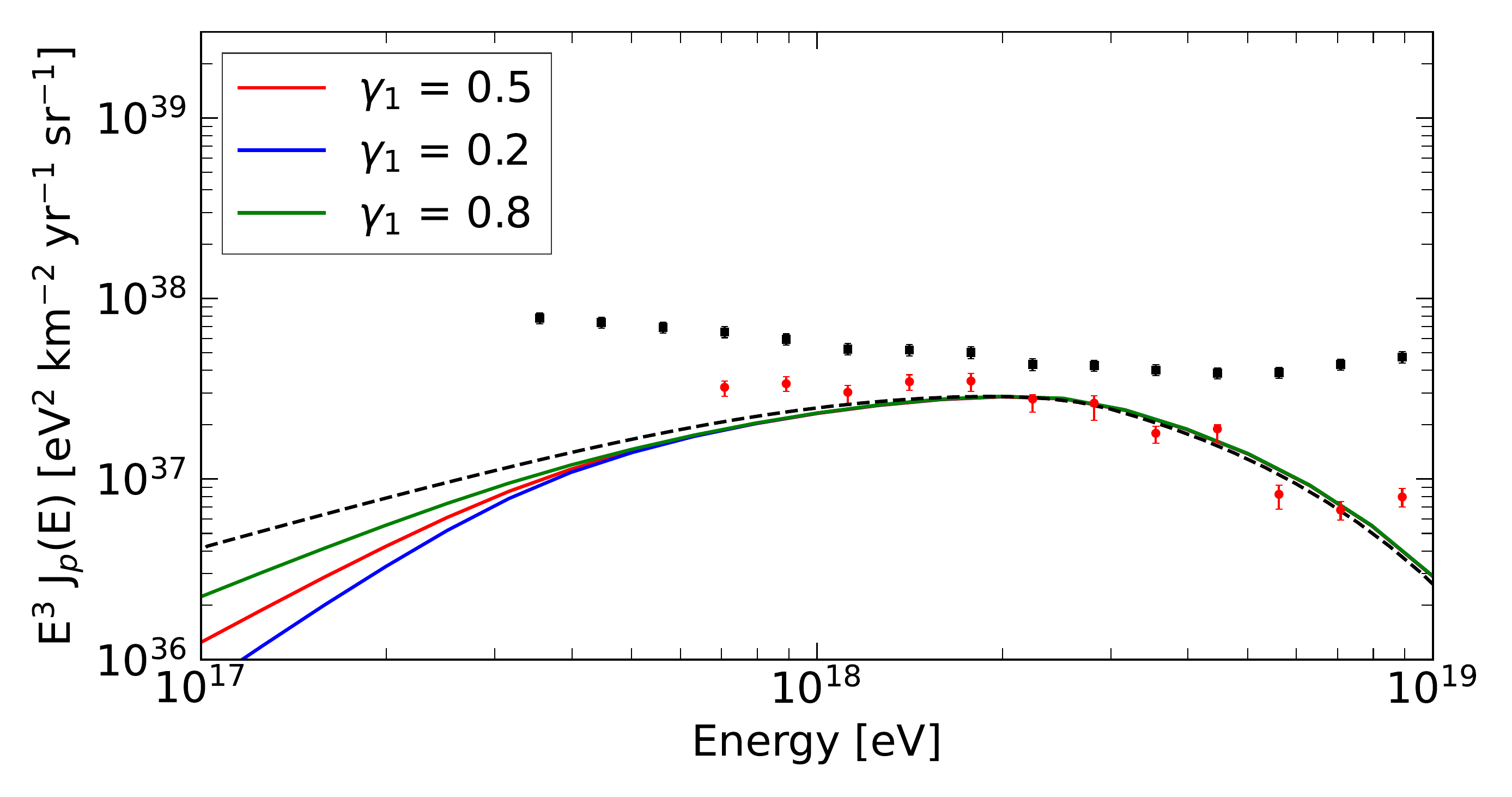}
    \caption{\label{fig:BPL_spectrum_slope}}
    \end{subfigure}
    \hfill
    \begin{subfigure}{0.48\textwidth}
    \includegraphics[width=\linewidth]{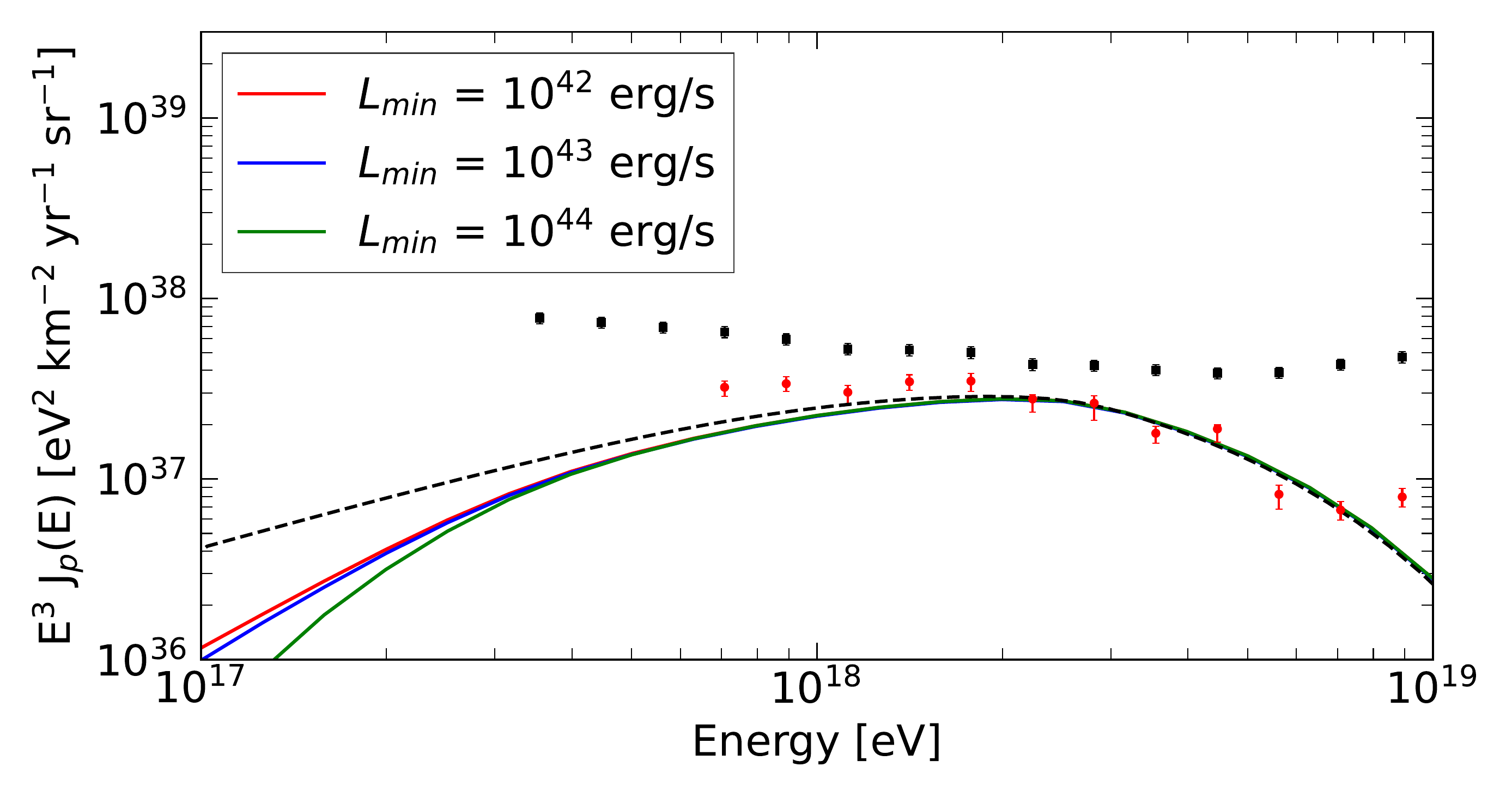}
    \caption{\label{fig:BPL_spectrum_minimum}}
    \end{subfigure}
\caption{Cosmic rays intensity at Earth. Top panel: different positioning of $\mathcal{L}_*$, with $\gamma_1 = 0.5$. Central panel: different slope, with the positioning of the break fixed at $\mathcal{L}_* = 10^{45}$ erg/s. Bottom panel: different choices of $\mathcal{L}_{\rm min}$, with $\mathcal{L}_* = 10^{45}$ erg/s and $\gamma_1 = 0.5$. $\gamma_2 = 2.5$ in all the cases. The all particle spectrum~\cite{abreu2021energy, PAO_AllPart_PhysRevD} measured by Auger and the proton fraction~\cite{XmaxPRD2014, Hfrac2014PRD, Xmax2016PRB} inferred through the EPOS-LHC hadronic interaction model~\cite{pierog2015epos} are shown with black squares and red circles, respectively. As a comparison, the same result is shown for an injection $Q_{src}(E) \propto E^{-2}$ (black dashed).\label{fig:figure4}}
\end{figure}

\begin{figure}[t]
    \centering
    \begin{subfigure}{0.48\textwidth}
    \includegraphics[width=\linewidth]{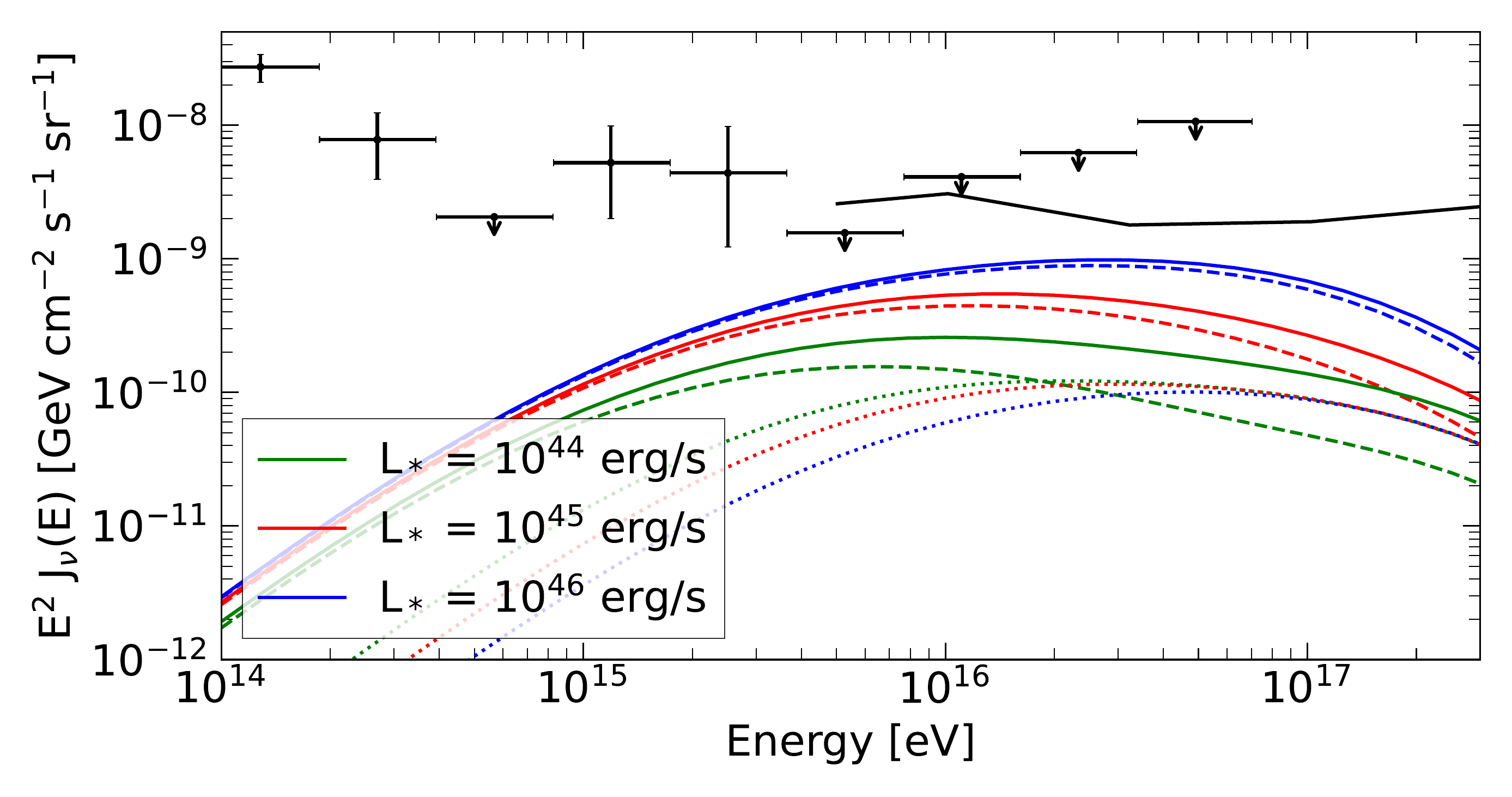}
    \caption{\label{fig:BPL_neutrino_lum}}
    \end{subfigure}
    \hfill
    \begin{subfigure}{0.48\textwidth}
    \includegraphics[width=\linewidth]{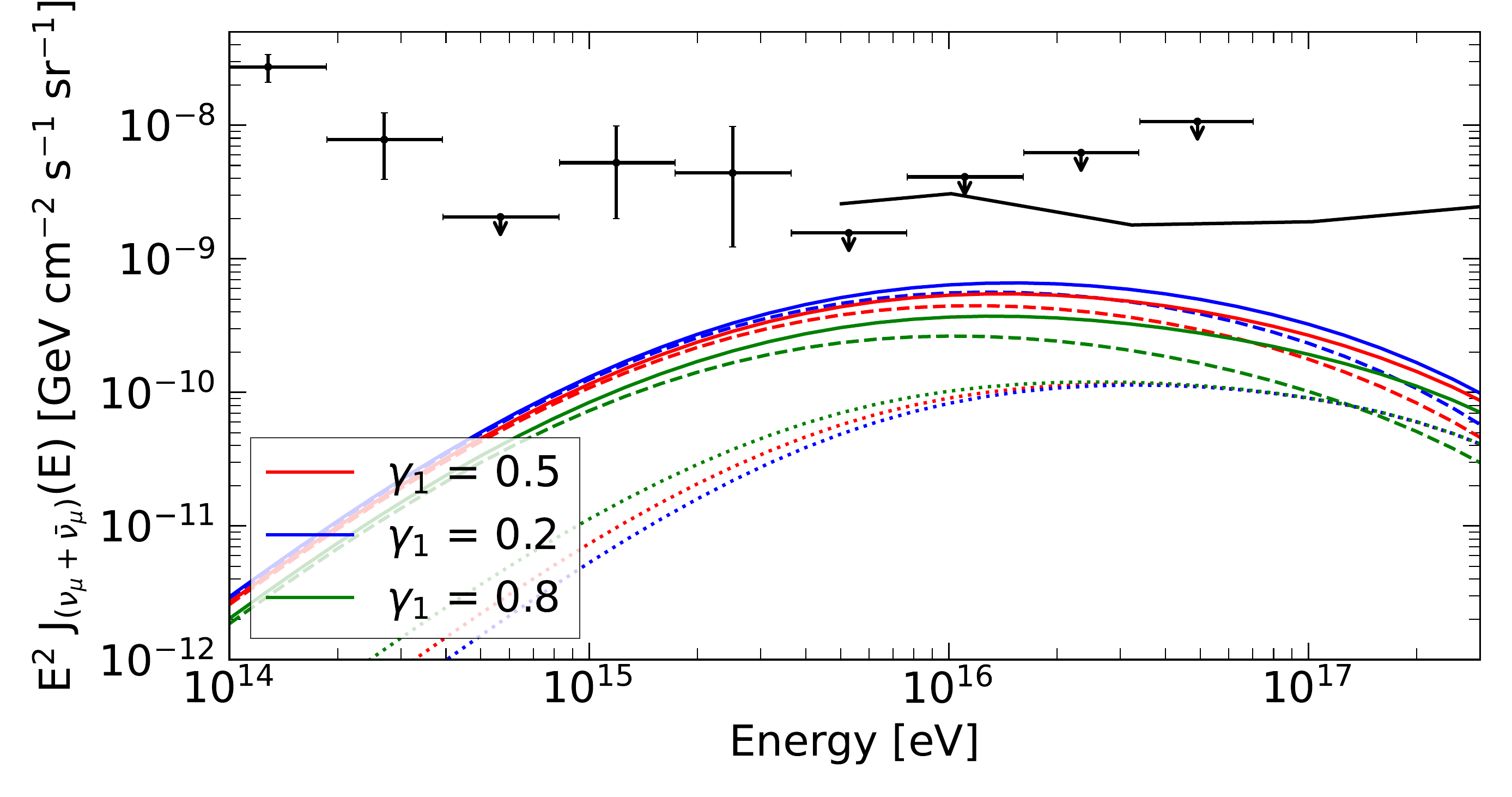}
    \caption{\label{fig:BPL_neutrino_slope}}
    \end{subfigure}
    \hfill
    \begin{subfigure}{0.48\textwidth}
    \includegraphics[width=\linewidth]{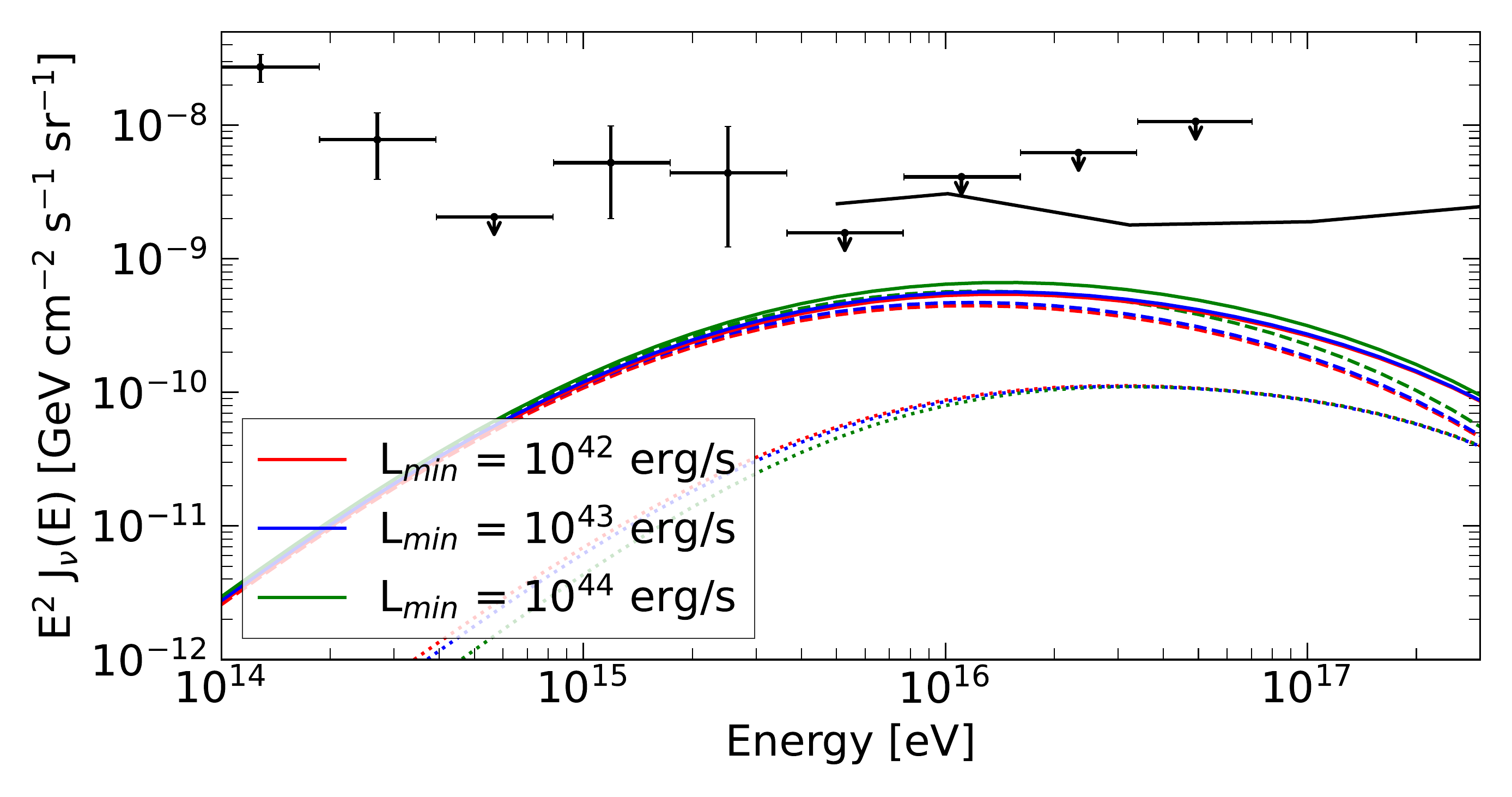}
    \caption{\label{fig:BPL_neutrino_minimum}}
    \end{subfigure}
    \caption{Total diffuse neutrino flux at Earth (solid), coming from confined protons (dashed lines) and diffused ones (dotted lines). Top panel: different positioning of $\mathcal{L}_*$, with $\gamma_1 = 0.5$. Central panel: different slope, with the positioning of the break fixed at $\mathcal{L}_* = 10^{45}$ erg/s. Bottom panel: different choices of $\mathcal{L}_{\rm min}$, with $\mathcal{L}_* = 10^{45}$ erg/s and $\gamma_1 = 0.5$. $\gamma_2 = 2.5$ in all the cases. Measurements (black dots) and upper limits (black line and arrows) obtained by the Ice Cube Observatory~\cite{Aartsen_2013, Kopper2017IceCubeLims, IceCube:2025ezc} are shown as a comparison.}\label{fig:figure5}
\end{figure}

\subsection{Comparison with observations}

The flux of CR protons resulting from sources distributed with a luminosity function is shown in Figure~\ref{fig:figure4}, together with the all-particle spectrum measured by Auger~\citep{abreu2021energy, PAO_AllPart_PhysRevD} and the proton spectrum obtained by adopting the mass fraction computed in~\citet{Xmax2016PRB, Hfrac2014PRD, XmaxPRD2014} using EPOS-LHC~\citep{pierog2015epos}. The parameters defining the luminosity function for all the tested cases are listed in Table~\ref{tab:tabella}.

The three panels refer to the same cases illustrated in Figure~\ref{fig:figure3}, and the normalization constant of the LF is chosen not to overshoot the proton spectrum measured by the Auger in the ankle region. 
As can be inferred from the comparison with an injection spectrum of E$^{-2}$ (black-dashed), the confinement mechanism we have proposed is not necessary to explain the fraction of protons around 10$^{18}$ eV, whereas it may be relevant for the cosmic ray spectrum above the ankle, where the inferred mass composition is increasingly heavier and the required spectral index at the escape from the source is unusually hard~\citep{AbdulHalim_2023}.  However, the possibility that the sources of the highest-energy cosmic rays also accelerate protons, although in a smaller fraction compared to heavier nuclei, cannot be excluded~\citep{muzio2024peterscycleendcosmic}. 
Moreover, in such a scenario, interactions within the confinement region lead to the production of secondary protons~\citep{UngerPRD}.

We then use the proton flux constrained by Auger data to derive an estimate of the contribution to the diffuse neutrino flux of these sources, provided the proposed mechanism is in place.
In doing so, we assume that the maximum proton content allowed for the UHECR population that dominates above the ankle is restricted by the proton fraction measured by Auger, as in Figure~\ref{fig:figure4}.

The two contributions, $J_\nu^\text{cosmo}$ and $J_\nu^\text{conf}$, are shown in Figure~\ref{fig:figure5} for the same cases illustrated in Figure~\ref{fig:figure4}, with the same choice of the normalization constant. The neutrino flux produced by confined protons exceeds the cosmogenic one in the (1-100) PeV energy range. However, the total neutrino flux remains compatible with the upper limits posed by the Ice Cube experiment~\citep{Kopper2017IceCubeLims, IceCube:2025ezc} even for the maximal choice about the proton flux, as discussed before.

A detailed study and comparison with the spectrum and mass composition as measured by Auger, including the nuclei and the related interactions inside the confinement environment, will be fully addressed in a forthcoming work.

\section{Discussion and conclusions}
\label{sec:discuss}

The Auger data on the UHECR spectrum and mass composition unequivocally indicate that the dominant population of sources above the ankle~\citep{AbdulHalim_2023} must inject a spectrum with a remarkably hard shape, and maximum energy in the range of a few EeV. Here, the term sources refers to astrophysical objects that may either host the real accelerators or act as storage environments for cosmic rays. This distinction is well illustrated by clusters of galaxies, which have long been recognized as natural reservoirs for cosmic rays over cosmological timescales~\citep{Berezinsky_1997, Ensslin1997}. Within such clusters, cosmic rays may be accelerated either by shocks in the intracluster medium or by individual galaxies. For cosmic rays with energies \( \lesssim \) PeV, the confinement time often exceeds the age of the universe, leading to a natural low-energy suppression that effectively hardens the spectrum. 
The Auger data suggest that such hardening likely occurs in the \( \lesssim \) EeV energy range.

This spectral hardening may also result from pronounced confinement processes similar to those inferred in galaxy clusters~\citep{Berezinsky_1997, Ensslin1997}. For instance, a combination of confinement and energy losses in specific sources~\citep{UngerPRD, Muzio2022PRD} or the magnetic horizon effect $-$ where strong extragalactic magnetic fields create a low energy cutoff~\citep{Aloisio_2004, MollerachJCAP, halim2024impact} $-$ can mimic this phenomenon. However, the latter explanation requires relatively large magnetic fields that may only be marginally consistent with observational constraints, such as Faraday rotation measures, if such fields are ubiquitous.

We propose an alternative mechanism for this hardening: self-confinement driven by the excitation of plasma instabilities near powerful UHECR sources. When the energy density of escaping UHECRs exceeds the background magnetic field energy density, self-generated turbulence can significantly alter particle transport. In this work, we extend previous studies~\citep{Blasi2015prl} by exploring the environmental conditions under which self-confinement occurs at EeV energies. We calculate the resulting UHECR proton spectrum and diffuse neutrino flux in this framework.

Our results confirm that the scattering of UHECRs on self-generated magnetic perturbations leads to confinement timescales exceeding the source age for particles with \( \lesssim \)~EeV energies. This effect imprints a low-energy suppression in the UHECR spectrum at Earth, influenced by the luminosity function of the sources.

The mechanism is particularly efficient if the source luminosity is of the order of  \( \sim 10^{45} \)~erg/s and if the pre-existing IGMF is in the range \( B_0 \sim 0.1-1 \) nG, within a region of \( \sim 10 \) Mpc around the source. Notably, the non-resonant instability may still grow for lower values of \( B_0 \), provided that particles down to \( \sim \) PeV contribute to the current (see Eq.~\eqref{eq:Blower}). Such conditions could naturally arise in environments like galaxy clusters, where the interplay between confinement and turbulence is prominent.

It is worth noticing that most potential UHECR accelerators, such as AGNs~\citep{Pimbblet2012} or sources in starburst galaxies~\citep{Paccagnella_2017}, are indeed expected to reside within galaxy clusters. These regions are more likely to be magnetized due to the filamentary structure of the large-scale universe~\citep{VernstromMNRAS, Carretti2022}.

We calculated the UHECR proton flux from a cosmological population of sources with a typical luminosity function and compared our predictions with the Auger proton fraction~\citep{abreu2021energy, PAO_AllPart_PhysRevD, XmaxPRD2014, Hfrac2014PRD, Xmax2016PRB} interpreted using the EPOS-LHC hadronic interaction model~\citep{pierog2015epos}. Our results predict an excess of neutrino production in the near-source regions due to self-confinement of UHECRs with \( E \lesssim 1 \) EeV, compared to the cosmogenic neutrino flux from intergalactic space. However, the total neutrino flux remains consistent with current observational upper limits~\citep{Kopper2017IceCubeLims, IceCube:2025ezc}.

Finally, our model predicts that powerful UHECR sources should be surrounded by regions extending tens of Mpc, with turbulent magnetic fields reaching nG levels due to the saturation of the non-resonant instability. Observations of X-ray and radio emissions already suggest the presence of magnetized structures connecting galaxy clusters~\citep{VernstromMNRAS}, with average magnetic fields of 30–50 nG on scales 
\( \gtrsim 3 \)~Mpc. Faraday rotation measures, which are sensitive to the mean magnetic field direction, impose upper limits on large-scale fields at the level of a few nG~\citep{OSullivan2020}. However, these limits scale linearly with the assumed electron density, implying that magnetic fields in overdense regions, such as filaments, could be at least an order of magnitude larger.

The impact of amplified magnetic fields on Faraday rotation remains challenging to quantify, as it depends on the source luminosity function. The average rotation measure (RM) can be estimated as RM~\( \sim f^{1/2} n_e \langle \delta B \rangle \)~\citep{Amaral2021}, where \(\langle \delta B \rangle \) is the average amplified field strength weighted by the source luminosity function, \( n_e \) is the mean electron density and \( f \) is the filling factor, determined by the product of the source density and the volume of a single flux tube. 

Assuming the same electron density as in~\citet{Amaral2021}, $n_e \simeq 10^{-5}$ cm$^{-3}$, we estimate RM~\( \lesssim 10^{-9} \mu\)G cm\(^{-3}\), roughly an order of magnitude smaller than the limits of the residual rotation measure limits reported in~\citet{Amaral2021}. Values of the averaged amplified field, filling factor, and RM are reported in table~\ref{tab:tabella} for all the tested cases.

\begin{acknowledgements}

This work was partially funded by the European Union - NextGenerationEU under the MUR National Innovation Ecosystem grant  ECS00000041 - VITALITY/ASTRA - CUP D13C21000430001. The work of PB was also partially funded by the European Union - Next Generation EU, through PRIN-MUR 2022TJW4EJ.

\end{acknowledgements}

\begin{center}
\begin{table}
\captionof{table}{Values of the normalization constant $A$ of the tested luminosity functions, the resulting density of sources able to produce the NRSI $n(\mathcal{L}_\text{min})$, the averaged amplified magnetic field $\left< B \right>$, the volume fraction occupied by these sources $f$, and the estimate of the effect on the Faraday Rotation from the magnetized cocoons, RM. The luminosity functions extends up to $\mathcal{L}_{high} = 10^{48}$ erg/s, with a faint end slope $\gamma_2 = 2.5$, in all cases. \label{tab:tabella}}
\begin{tabular}{ |c|c|c|c| }
\hline
\multicolumn{4}{|c|}{ Different $\mathcal{L}_*$ - $\mathcal{L}_{low} = 10^{40} \, \rm erg/s$ - $\gamma_1 = 0.5$} \\
\hline
$\mathcal{L}_{*}$ \, $ \left[ \rm erg/s \right]$ &  $10^{44}$ & $10^{45}$ & $10^{46}$ \\
\hline
$A$ \, $ \left[ {\rm Mpc}^{-3} \right]$ & $2.2 \times 10^{-6}$ & $2.2 \times 10^{-7}$ & $2.2 \times 10^{-8}$ \\
\hline
$n(\mathcal{L}_{min})$ \, $ \left[ {\rm Mpc}^{-3} \right]$ & $1.3 \times 10^{-5}$ & $4.7 \times 10^{-6}$ & $1.5 \times 10^{-6}$ \\
\hline
$\left< B \right>$ \, $ \left[ \rm nG \right]$ & 2.4 & 3.4 & 4.4 \\
\hline
$f$ & $8.3 \times 10^{-4}$ & $2.9 \times 10^{-4}$ & $9.5 \times 10^{-5}$ \\
\hline
$\rm RM$ \, $ \left[ \mu {\rm G \, cm}^{-3} \right]$ & $6.8 \times 10^{-10}$ & $5.8 \times 10^{-10}$ & $4.3 \times 10^{-10}$ \\
\hline \hline
\multicolumn{4}{|c|}{ Different $\gamma_1$ - $\mathcal{L}_* = 10^{45} \, \rm erg/s$ - $\mathcal{L}_{low} = 10^{40} \, \rm erg/s$ } \\
\hline
$\gamma_{1}$ & 0.2 & 0.5 & 0.8 \\
\hline
$A$ \, $ \left[ {\rm Mpc}^{-3} \right]$ & $3.2 \times 10^{-7}$ & $2.2 \times 10^{-7}$ & $1.1 \times 10^{-7}$ \\
\hline
$n(\mathcal{L}_{min})$ \, $ \left[ {\rm Mpc}^{-3} \right]$ & $1.8 \times 10^{-6}$ & $4.7 \times 10^{-6}$ & $1.0 \times 10^{-5}$ \\
\hline
$\left< B \right>$ \, $ \left[ \rm nG \right]$ & 5.7 & 3.4 & 2.3 \\
\hline
$f$ & $1.1 \times 10^{-4}$ & $2.9 \times 10^{-4}$ & $6.4 \times 10^{-4}$ \\
\hline
$\rm RM$ \, $ \left[ \mu {\rm G \, cm}^{-3} \right]$ & $6.1 \times 10^{-10}$ & $5.8 \times 10^{-10}$ & $5.8 \times 10^{-10}$ \\
\hline \hline
\multicolumn{4}{|c|}{ Different $\mathcal{L}_{low}$ - $\mathcal{L}_* = 10^{45} \, \rm erg/s$ - $\gamma_1 = 0.5$} \\
\hline
$\mathcal{L}_{low}$ \, $ \left[ \rm erg/s \right]$ & $10^{42}$ & $10^{43}$ & $10^{44}$ \\
\hline
$A$ \, $ \left[ {\rm Mpc}^{-3} \right]$ & $2.2 \times 10^{-7}$ & $2.3 \times 10^{-7}$ & $3.0 \times 10^{-7}$ \\
\hline
$n(\mathcal{L}_{min})$ \, $ \left[ {\rm Mpc}^{-3} \right]$ & $5.9 \times 10^{-6}$ & $1.8 \times 10^{-6}$ & $5.4 \times 10^{-7}$\\
\hline
$\left< B \right>$ \, $ \left[ \rm nG \right]$ & 2.7 & 6.6 & 14.2 \\
\hline
$f$ & $3.7 \times 10^{-4}$ & $1.1 \times 10^{-4}$ & $3.4 \times 10^{-5}$ \\
\hline
$\rm RM$ \, $ \left[ \mu {\rm G \, cm}^{-3} \right]$ & $5.2 \times 10^{-10}$ & $7.0 \times 10^{-10}$ & $8.2 \times 10^{-10}$ \\
\hline

\end{tabular}
\end{table}
\end{center}

\bibliographystyle{aa}
\bibliography{Cermenati2024}

\begin{thebibliography}{58}
\expandafter\ifx\csname natexlab\endcsname\relax\def\natexlab#1{#1}\fi

\bibitem[{Aab {et~al.}(2016)Aab, Abreu, Aglietta, Ahn, {Al Samarai}, Albuquerque, Allekotte, Allison, Almela, {Alvarez Castillo}, Alvarez-Muñiz, Ambrosio, Anastasi, Anchordoqui, Andrada, Andringa, Aramo, Arqueros, Arsene, Asorey, Assis, Aublin, Avila, Badescu, Balaceanu, Baus, Beatty, Becker, Bellido, Berat, Bertaina, Biermann, Billoir, Biteau, Blaess, Blanco, Blazek, Bleve, Boháčová, Boncioli, Bonifazi, Borodai, Botti, Brack, Brancus, Bretz, Bridgeman, Briechle, Buchholz, Bueno, Buitink, Buscemi, Caballero-Mora, Caccianiga, Caccianiga, Cancio, Canfora, Caramete, Caruso, Castellina, Cataldi, Cazon, Cester, Chavez, Chiavassa, Chinellato, Chudoba, Clay, Colalillo, Coleman, Collica, Coluccia, Conceição, Contreras, Cooper, Coutu, Covault, Cronin, Dallier, D'Amico, Daniel, Dasso, Daumiller, Dawson, {de Almeida}, {de Jong}, {De Mauro}, {de Mello Neto}, {De Mitri}, {de Oliveira}, {de Souza}, Debatin, {del Peral}, Deligny, {Di Giulio}, {Di Matteo}, {Díaz Castro}, Diogo, Dobrigkeit, D'Olivo, Dorofeev, {dos
  Anjos}, Dova, Dundovic, Ebr, Engel, Erdmann, Erfani, Escobar, Espadanal, Etchegoyen, Falcke, Fang, Farrar, Fauth, Fazzini, Fick, Figueira, Filevich, Filipčič, Fratu, Freire, Fujii, Fuster, García, Garcia-Pinto, Gaté, Gemmeke, Gherghel-Lascu, Ghia, Giaccari, Giammarchi, Giller, Głas, Glaser, Glass, Golup, {Gómez Berisso}, {Gómez Vitale}, González, Gookin, Gordon, Gorgi, Gorham, Gouffon, Grillo, Grubb, Guarino, Guedes, Hampel, Hansen, Harari, Harrison, Harton, Hasankiadeh, Haungs, Hebbeker, Heck, Heimann, Herve, Hill, Hojvat, Holt, Homola, Hörandel, Horvath, Hrabovský, Huege, Hulsman, Insolia, Isar, Jandt, Jansen, Johnsen, Josebachuili, Kääpä, Kambeitz, Kampert, Kasper, Katkov, Keilhauer, Kemp, Kieckhafer, Klages, Kleifges, Kleinfeller, Krause, Krohm, Kuempel, {Kukec Mezek}, Kunka, {Kuotb Awad}, LaHurd, Latronico, Lauscher, Lautridou, Lebrun, Legumina, {Leigui de Oliveira}, Letessier-Selvon, Lhenry-Yvon, Link, Lopes, López, {López Casado}, Luce, Lucero, Malacari, Mallamaci, Mandat, Mantsch,
  Mariazzi, Mariş, Marsella, Martello, Martinez, {Martínez Bravo}, {Masías Meza}, Mathes, Mathys, Matthews, Matthews, Matthiae, Mayotte, Mazur, Medina, Medina-Tanco, Melo, Menshikov, Messina, Micheletti, Middendorf, Minaya, Miramonti, Mitrica, Mockler, Molina-Bueno, Mollerach, Montanet, Morello, Mostafá, Müller, Muller, Müller, Naranjo, Navas, Nellen, Neuser, Nguyen, Niculescu-Oglinzanu, Niechciol, Niemietz, Niggemann, Nitz, Nosek, Novotny, Nožka, Núñez, Ochilo, Oikonomou, Olinto, {Pakk Selmi-Dei}, Palatka, Pallotta, Papenbreer, Parente, Parra, Paul, Pech, Pedreira, Pȩkala, Pelayo, Peña-Rodriguez, Pereira, Perrone, Peters, Petrera, Phuntsok, Piegaia, Pierog, Pieroni, Pimenta, Pirronello, Platino, Plum, Porowski, Prado, Privitera, Prouza, Quel, Querchfeld, Quinn, Ramos-Pollant, Rautenberg, Ravel, Ravignani, Reinert, Revenu, Ridky, Risse, Ristori, Rizi, {Rodrigues de Carvalho}, {Rodriguez Fernandez}, {Rodriguez Rojo}, Rodríguez-Frías, Rogozin, Rosado, Roth, Roulet, Rovero, Saffi, Saftoiu, Salazar,
  Saleh, {Salesa Greus}, Salina, {Sanabria Gomez}, Sánchez, Sanchez-Lucas, Santos, Santos, Sarazin, Sarkar, Sarmento, Sarmiento-Cano, Sato, Scarso, Schauer, Scherini, Schieler, Schmidt, Scholten, Schovánek, Schröder, Schulz, Schulz, Schumacher, Sciutto, Segreto, Settimo, Shadkam, Shellard, Sigl, Silli, Sima, Śmiałkowski, Šmída, Snow, Sommers, Sonntag, Sorokin, Squartini, Stanca, Stanič, Stasielak, Strafella, Suarez, {Suarez Durán}, Sudholz, Suomijärvi, Supanitsky, Sutherland, Swain, Szadkowski, Taborda, Tapia, Tepe, Theodoro, Timmermans, {Todero Peixoto}, Tomankova, Tomé, Tonachini, {Torralba Elipe}, {Torres Machado}, Torri, Travnicek, Trini, Ulrich, Unger, Urban, Valbuena-Delgado, {Valdés Galicia}, Valiño, Valore, {van Aar}, {van Bodegom}, {van den Berg}, {van Vliet}, Varela, {Vargas Cárdenas}, Varner, Vázquez, Vázquez, Veberič, Verzi, Vicha, Villaseñor, Vorobiov, Wahlberg, Wainberg, Walz, Watson, Weber, Weindl, Wiencke, Wilczyński, Winchen, Wittkowski, Wundheiler, Wykes, Yang, Yelos,
  Younk, Yushkov, Zas, Zavrtanik, Zavrtanik, Zepeda, Zimmermann, Ziolkowski, Zong, \& Zuccarello}]{Xmax2016PRB}
Aab, A., Abreu, P., Aglietta, M., {et~al.} 2016, Physics Letters B, 762, 288

\bibitem[{Aab {et~al.}(2014)Aab, Abreu, Aglietta, Ahn, Al~Samarai, Albuquerque, Allekotte, Allen, Allison, Almela, Alvarez~Castillo, Alvarez-Mu\~niz, Alves~Batista, Ambrosio, Aminaei, Anchordoqui, Andringa, Aramo, Aranda, Arqueros, Asorey, Assis, Aublin, Ave, Avenier, Avila, Awal, Badescu, Barber, B\"auml, Baus, Beatty, Becker, Bellido, Berat, Bertania, Bertou, Biermann, Billoir, Blaess, Blanco, Bleve, Bl\"umer, Boh\'a\ifmmode~\check{c}\else \v{c}\fi{}ov\'a, Boncioli, Bonifazi, Bonino, Borodai, Brack, Brancus, Bridgeman, Brogueira, Brown, Buchholz, Bueno, Buitink, Buscemi, Caballero-Mora, Caccianiga, Caccianiga, Candusso, Caramete, Caruso, Castellina, Cataldi, Cazon, Cester, Chavez, Chiavassa, Chinellato, Chudoba, Cilmo, Clay, Cocciolo, Colalillo, Coleman, Collica, Coluccia, Concei\ifmmode \mbox{\c{c}}\else~\c{c}\fi{}\ ao, Contreras, Cooper, Cordier, Coutu, Covault, Cronin, Curutiu, Dallier, Daniel, Dasso, Daumiller, Dawson, de~Almeida, De~Domenico, de~Jong, de~Mello~Neto, De~Mitri, de~Oliveira, de~Souza, del
  Peral, Deligny, Dembinski, Dhital, Di~Giulio, Di~Matteo, Diaz, D\'{\i}az~Castro, Diogo, Dobrigkeit, Docters, D'Olivo, Dorofeev, Dorosti~Hasankiadeh, Dova, Ebr, Engel, Erdmann, Erfani, Escobar, Espadanal, Etchegoyen, Facal San~Luis, Falcke, Fang, Farrar, Fauth, Fazzini, Ferguson, Fernandes, Fick, Figueira, Filevich, Filip\ifmmode \check{c}\else \v{c}\fi{}i\ifmmode~\check{c}\else \v{c}\fi{}, Fox, Fratu, Fr\"ohlich, Fuchs, Fuji, Gaior, Garc\'{\i}a, Garcia~Roca, Garcia-Gamez, Garcia-Pinto, Garilli, Gascon~Bravo, Gate, Gemmeke, Ghia, Giaccari, Giammarchi, Giller, Glaser, Glass, G\'omez~Berisso, G\'omez~Vitale, Gon\ifmmode~\mbox{\c{c}}\else \c{c}\fi{}alves, Gonzalez, Gonz\'alez, Gookin, Gordon, Gorgi, Gorham, Gouffon, Grebe, Griffith, Grillo, Grubb, Guarino, Guedes, Hampel, Hansen, Harari, Harrison, Hartmann, Harton, Haungs, Hebbeker, Heck, Heimann, Herve, Hill, Hojvat, Hollon, Holt, Homola, H\"orandel, Horvath, Hrabovsk\'y, Huber, Huege, Insolia, Isar, Jandt, Jansen, Jarne, Josebachuili, K\"a\"ap\"a, Kambeitz,
  Kampert, Kasper, Katkov, K\'egl, Keilhauer, Keivani, Kemp, Kieckhafer, Klages, Kleifges, Kleinfeller, Krause, Krohm, Kr\"omer, Kruppke-Hansen, Kuempel, Kunka, LaHurd, Latronico, Lauer, Lauscher, Lautridou, Le~Coz, Le\~ao, Lebrun, Lebrun, Leigui~de Oliveira, Letessier-Selvon, Lhenry-Yvon, Link, L\'opez, Lopez~Ag\"uera, Louedec, Lozano~Bahilo, Lu, Lucero, Ludwig, Malacari, Maldera, Mallamaci, Maller, Mandat, Mantsch, Mariazzi, Marin, Mari\ifmmode~\mbox{\c{s}}\else \c{s}\fi{}, Marsella, Martello, Martin, Martinez, Mart\'{\i}nez~Bravo, Martraire, Mas\'{\i}as~Meza, Mathes, Mathys, Matthews, Matthews, Matthiae, Maurel, Maurizio, Mayotte, Mazur, Medina, Medina-Tanco, Meissner, Melissas, Melo, Menshikov, Messina, Meyhandan, Mi\ifmmode \acute{c}\else \'{c}\fi{}anovi\ifmmode~\acute{c}\else \'{c}\fi{}, Micheletti, Middendorf, Minaya, Miramonti, Mitrica, Molina-Bueno, Mollerach, Monasor, Monnier~Ragaigne, Montanet, Morello, Mostaf\'a, Moura, Muller, M\"uller, M\"uller, M\"unchmeyer, Mussa, Navarra, Navas, Necesal,
  Nellen, Nelles, Neuser, Nguyen, Niechciol, Niemietz, Niggemann, Nitz, Nosek, Novotny, No\ifmmode~\check{z}\else \v{z}\fi{}ka, Ochilo, Olinto, Oliveira, Pacheco, Pakk Selmi-Dei, Palatka, Pallotta, Palmieri, Papenbreer, Parente, Parra, Paul, Pech, P\ifmmode~\mbox{\c{e}}\else \c{e}\fi{}kala, Pelayo, Pepe, Perrone, Petermann, Peters, Petrera, Petrov, Phuntsok, Piegaia, Pierog, Pieroni, Pimenta, Pirronello, Platino, Plum, Porcelli, Porowski, Prado, Privitera, Prouza, Purrello, Quel, Querchfeld, Quinn, Rautenberg, Ravel, Ravignani, Revenu, Ridky, Riggi, Risse, Ristori, Rizi, Rodrigues~de Carvalho, Rodriguez~Cabo, Rodriguez~Fernandez, Rodriguez~Rojo, Rodr\'{\i}guez-Fr\'{\i}as, Rogozin, Ros, Rosado, Rossler, Roth, Roulet, Rovero, Saffi, Saftoiu, Salamida, Salazar, Saleh, Salesa~Greus, Salina, S\'anchez, Sanchez-Lucas, Santo, Santos, Santos, Sarazin, Sarkar, Sarmento, Sato, Scharf, Scherini, Schieler, Schiffer, Schmidt, Scholten, Schoorlemmer, Schov\'anek, Schulz, Schulz, Schumacher, Sciutto, Segreto, Settimo,
  Shadkam, Shellard, Sidelnik, Sigl, Sima, \ifmmode~\acute{S}\else \'{S}\fi{}mia\l{}kowski, \ifmmode~\check{S}\else \v{S}\fi{}m\'{\i}da, Snow, Sommers, Sorokin, Squartini, Srivastava, Stani\ifmmode~\check{c}\else \v{c}\fi{}, Stapleton, Stasielak, Stephan, Stutz, Suarez, Suomij\"arvi, Supanitsky, Sutherland, Swain, Szadkowski, Szuba, Taborda, Tapia, Tartare, Tepe, Theodoro, Timmermans, Todero~Peixoto, Toma, Tomankova, Tom\'e, Tonachini, Torralba~Elipe, Torres~Machado, Travnicek, Trovato, Tueros, Ulrich, Unger, Urban, Vald\'es~Galicia, Vali\~no, Valore, van Aar, van Bodegom, van~den Berg, van Velzen, van Vliet, Varela, Vargas~C\'ardenas, Varner, V\'azquez, V\'azquez, Veberi\ifmmode~\check{c}\else \v{c}\fi{}, Verzi, Vicha, Videla, Villase\~nor, Vlcek, Vorobiov, Wahlberg, Wainberg, Walz, Watson, Weber, Weidenhaupt, Weindl, Werner, Widom, Wiencke, Wilczy\ifmmode~\acute{n}\else \'{n}\fi{}ska, Wilczy\ifmmode~\acute{n}\else \'{n}\fi{}ski, Will, Williams, Winchen, Wittkowski, Wundheiler, Wykes, Yamamoto, Yapici, Yuan,
  Yushkov, Zamorano, Zas, Zavrtanik, Zavrtanik, Zaw, Zepeda, Zhou, Zhu, Zimbres~Silva, Ziolkowski, \& Zuccarello}]{XmaxPRD2014}
Aab, A., Abreu, P., Aglietta, M., {et~al.} 2014, Phys. Rev. D, 90, 122006

\bibitem[{{Aab} {et~al.}(2014){Aab}, {Abreu}, {Aglietta}, {Ahn}, {Al Samarai}, {Albuquerque}, {Allekotte}, {Allen}, {Allison}, {Almela}, \& et~al.}]{Hfrac2014PRD}
{Aab}, A., {Abreu}, P., {Aglietta}, M., {et~al.} 2014, \prd, 90, 122005

\bibitem[{Aab {et~al.}(2020)Aab, Abreu, Aglietta, Albury, Allekotte, Almela, Alvarez~Castillo, Alvarez-Mu\~niz, Alves~Batista, Anastasi, Anchordoqui, Andrada, Andringa, Aramo, Ara\'ujo~Ferreira, Asorey, Assis, Avila, Badescu, Bakalova, Balaceanu, Barbato, Barreira~Luz, Becker, Bellido, Berat, Bertaina, Bertou, Biermann, Bister, Biteau, Blanco, Blazek, Bleve, Boh\'a\ifmmode~\check{c}\else \v{c}\fi{}ov\'a, Boncioli, Bonifazi, Bonneau~Arbeletche, Borodai, Botti, Brack, Bretz, Briechle, Buchholz, Bueno, Buitink, Buscemi, Caballero-Mora, Caccianiga, Calcagni, Cancio, Canfora, Caracas, Carceller, Caruso, Castellina, Catalani, Cataldi, Cazon, Cerda, Chinellato, Choi, Chudoba, Chytka, Clay, Cobos~Cerutti, Colalillo, Coleman, Coluccia, Concei\ifmmode \mbox{\c{c}}\else~\c{c}\fi{}\ ao, Condorelli, Consolati, Contreras, Convenga, Covault, Dasso, Daumiller, Dawson, Day, de~Almeida, de~Jes\'us, de~Jong, De~Mauro, de~Mello~Neto, De~Mitri, de~Oliveira, de~Oliveira~Franco, de~Souza, De~Vito, Debatin, del R\'{\i}o, Deligny,
  Dembinski, Dhital, Di~Giulio, Di~Matteo, D\'{\i}az~Castro, Dobrigkeit, D'Olivo, Dorosti, dos Anjos, Dova, Ebr, Engel, Epicoco, Erdmann, Escobar, Etchegoyen, Falcke, Farmer, Farrar, Fauth, Fazzini, Feldbusch, Fenu, Fick, Figueira, Filip\ifmmode \check{c}\else \v{c}\fi{}i\ifmmode~\check{c}\else \v{c}\fi{}, Fodran, Freire, Fujii, Fuster, Galea, Galelli, Garc\'{\i}a, Garcia~Vegas, Gemmeke, Gesualdi, Gherghel-Lascu, Ghia, Giaccari, Giammarchi, Giller, Glombitza, Gobbi, Gollan, Golup, G\'omez~Berisso, G\'omez~Vitale, Gongora, Gonz\'alez, Goos, G\'ora, Gorgi, Gottowik, Grubb, Guarino, Guedes, Guido, Hahn, Halliday, Hampel, Hansen, Harari, Harvey, Haungs, Hebbeker, Heck, Hill, Hojvat, H\"orandel, Horvath, Hrabovsk\'y, Huege, Hulsman, Insolia, Isar, Johnsen, Jurysek, K\"a\"ap\"a, Kampert, Keilhauer, Kemp, Klages, Kleifges, Kleinfeller, K\"opke, Kukec~Mezek, Lago, LaHurd, Lang, Leigui~de Oliveira, Lenok, Letessier-Selvon, Lhenry-Yvon, Lo~Presti, Lopes, L\'opez, Lorek, Luce, Lucero, Machado~Payeras, Malacari,
  Mancarella, Mandat, Manning, Manshanden, Mantsch, Marafico, Mariazzi, Mari\ifmmode~\mbox{\c{s}}\else \c{s}\fi{}, Marsella, Martello, Martinez, Mart\'{\i}nez~Bravo, Mastrodicasa, Mathes, Matthews, Matthiae, Mayotte, Mazur, Medina-Tanco, Melo, Menshikov, Merenda, Michal, Micheletti, Miramonti, Mockler, Mollerach, Montanet, Morello, Mostaf\'a, M\"uller, Muller, Mulrey, Mussa, Muzio, Namasaka, Nellen, Nguyen, Niculescu-Oglinzanu, Niechciol, Nitz, Nosek, Novotny, No\ifmmode~\check{z}\else \v{z}\fi{}ka, Nucita, N\'u\~nez, Palatka, Pallotta, Panetta, Papenbreer, Parente, Parra, Pech, Pedreira, P\ifmmode~\mbox{\c{e}}\else \c{e}\fi{}kala, Pelayo, Pe\~na Rodriguez, Perez~Armand, Perlin, Perrone, Peters, Petrera, Pierog, Pimenta, Pirronello, Platino, Pont, Pothast, Privitera, Prouza, Puyleart, Querchfeld, Rautenberg, Ravignani, Reininghaus, Ridky, Riehn, Risse, Ristori, Rizi, Rodrigues~de Carvalho, Rodriguez~Fernandez, Rodriguez~Rojo, Roncoroni, Roth, Roulet, Rovero, Ruehl, Saffi, Saftoiu, Salamida, Salazar, Salina,
  Sanabria~Gomez, S\'anchez, Santos, Santos, Sarazin, Sarmento, Sarmiento-Cano, Sato, Savina, Sch\"afer, Scherini, Schieler, Schimassek, Schimp, Schl\"uter, Schmidt, Scholten, Schov\'anek, Schr\"oder, Schr\"oder, Schulz, Sciutto, Scornavacche, Shellard, Sigl, Silli, Sima, \ifmmode~\check{S}\else \v{S}\fi{}m\'{\i}da, Sommers, Soriano, Souchard, Squartini, Stadelmaier, Stanca, Stani\ifmmode~\check{c}\else \v{c}\fi{}, Stasielak, Stassi, Streich, Su\'arez-Dur\'an, Sudholz, Suomij\"arvi, Supanitsky, \ifmmode~\check{S}\else \v{S}\fi{}up\'{\i}k, Szadkowski, Taboada, Tapia, Timmermans, Tkachenko, Tobiska, Todero~Peixoto, Tom\'e, Torralba~Elipe, Travaini, Travnicek, Trimarelli, Trini, Tueros, Ulrich, Unger, Urban, Vaclavek, Vacula, Vald\'es~Galicia, Vali\~no, Valore, van Vliet, Varela, Vargas~C\'ardenas, V\'asquez-Ram\'{\i}rez, Veberi\ifmmode~\check{c}\else \v{c}\fi{}, Ventura, Vergara~Quispe, Verzi, Vicha, Villase\~nor, Vink, Vorobiov, Wahlberg, Watson, Weber, Weindl, Wiencke, Wilczy\ifmmode~\acute{n}\else
  \'{n}\fi{}ski, Winchen, Wirtz, Wittkowski, Wundheiler, Yushkov, Zapparrata, Zas, Zavrtanik, Zavrtanik, Zehrer, Zepeda, Ziolkowski, \& Zuccarello}]{PAO_AllPart_PhysRevD}
Aab, A., Abreu, P., Aglietta, M., {et~al.} 2020, Phys. Rev. D, 102, 062005

\bibitem[{Aab {et~al.}(2017)Aab, Abreu, Aglietta, Samarai, Albuquerque, Allekotte, Almela, Castillo, Alvarez-Muñiz, Anastasi, Anchordoqui, Andrada, Andringa, Aramo, Arqueros, Arsene, Asorey, Assis, Aublin, Avila, Badescu, Balaceanu, Luz, Beatty, Becker, Bellido, Berat, Bertaina, Bertou, Biermann, Billoir, Biteau, Blaess, Blanco, Blazek, Bleve, Boháčová, Boncioli, Bonifazi, Borodai, Botti, Brack, Brancus, Bretz, Bridgeman, Briechle, Buchholz, Bueno, Buitink, Buscemi, Caballero-Mora, Caccianiga, Cancio, Canfora, Caramete, Caruso, Castellina, Cataldi, Cazon, Chavez, Chinellato, Chudoba, Clay, Colalillo, Coleman, Collica, Coluccia, Conceição, Contreras, Cooper, Coutu, Covault, Cronin, D'Amico, Daniel, Dasso, Daumiller, Dawson, de~Almeida, de~Jong, Mauro, de~Mello~Neto, Mitri, de~Oliveira, de~Souza, Debatin, Deligny, Giulio, di~Matteo, Castro, Diogo, Dobrigkeit, D'Olivo, Dorosti, dos Anjos, Dova, Dundovic, Ebr, Engel, Erdmann, Erfani, Escobar, Espadanal, Etchegoyen, Falcke, Farrar, Fauth, Fazzini, Fick,
  Figueira, Filipčič, Fratu, Freire, Fujii, Fuster, Gaior, García, Garcia-Pinto, Gaté, Gemmeke, Gherghel-Lascu, Ghia, Giaccari, Giammarchi, Giller, Głas, Glaser, Golup, Berisso, Vitale, González, Gorgi, Gorham, Grillo, Grubb, Guarino, Guedes, Hampel, Hansen, Harari, Harrison, Harton, Haungs, Hebbeker, Heck, Heimann, Herve, Hill, Hojvat, Holt, Homola, Hörandel, Horvath, Hrabovský, Huege, Hulsman, Insolia, Isar, Jandt, Jansen, Johnsen, Josebachuili, Kääpä, Kambeitz, Kampert, Katkov, Keilhauer, Kemp, Kemp, Kieckhafer, Klages, Kleifges, Kleinfeller, Krause, Krohm, Kuempel, Mezek, Kunka, Awad, LaHurd, Lauscher, Legumina, de~Oliveira, Letessier-Selvon, Lhenry-Yvon, Link, Lopes, López, Casado, Luce, Lucero, Malacari, Mallamaci, Mandat, Mantsch, Mariazzi, Mariš, Marsella, Martello, Martinez, Bravo, Meza, Mathes, Mathys, Matthews, Matthews, Matthiae, Mayotte, Mazur, Medina, Medina-Tanco, Melo, Menshikov, Micheletti, Middendorf, Minaya, Miramonti, Mitrica, Mockler, Mollerach, Montanet, Morello, Mostafá,
  Müller, Müller, Muller, Müller, Mussa, Naranjo, Nellen, Nguyen, Niculescu-Oglinzanu, Niechciol, Niemietz, Niggemann, Nitz, Nosek, Novotny, Nožka, Núñez, Ochilo, Oikonomou, Olinto, Palatka, Pallotta, Papenbreer, Parente, Parra, Paul, Pech, Pedreira, Pȩkala, Pelayo, Peña-Rodriguez, Pereira, Perlín, Perrone, Peters, Petrera, Phuntsok, Piegaia, Pierog, Pieroni, Pimenta, Pirronello, Platino, Plum, Porowski, Prado, Privitera, Prouza, Quel, Querchfeld, Quinn, Ramos-Pollan, Rautenberg, Ravignani, Revenu, Ridky, Risse, Ristori, Rizi, de~Carvalho, Fernandez, Rojo, Rogozin, Roncoroni, Roth, Roulet, Rovero, Ruehl, Saffi, Saftoiu, Salamida, Salazar, Saleh, Greus, Salina, Sánchez, Sanchez-Lucas, Santos, Santos, Sarazin, Sarmento, Sarmiento, Sato, Schauer, Scherini, Schieler, Schimp, Schmidt, Scholten, Schovánek, Schröder, Schulz, Schulz, Schumacher, Sciutto, Segreto, Settimo, Shadkam, Shellard, Sigl, Silli, Sima, Śmiałkowski, Šmída, Snow, Sommers, Sonntag, Sorokin, Squartini, Stanca, Stanič, Stasielak,
  Stassi, Strafella, Suarez, Durán, Sudholz, Suomijärvi, Supanitsky, Swain, Szadkowski, Taboada, Taborda, Tapia, Theodoro, Timmermans, Peixoto, Tomankova, Tomé, Elipe, Travnicek, Trini, Ulrich, Unger, Urban, Galicia, Valiño, Valore, van Aar, van Bodegom, van~den Berg, van Vliet, Varela, Cárdenas, Varner, Vázquez, Vázquez, Veberič, Quispe, Verzi, Vicha, Villaseñor, Vorobiov, Wahlberg, Wainberg, Walz, Watson, Weber, Weindl, Wiencke, Wilczyński, Winchen, Wirtz, Wittkowski, Wundheiler, Yang, Yelos, Yushkov, Zas, Zavrtanik, Zavrtanik, Zepeda, Zimmermann, Ziolkowski, Zong, \& Zong}]{Aab_2017JCAP}
Aab, A., Abreu, P., Aglietta, M., {et~al.} 2017, Journal of Cosmology and Astroparticle Physics, 2017, 038

\bibitem[{Aartsen {et~al.}(2013)Aartsen, Abbasi, Abdou, Ackermann, Adams, Aguilar, Ahlers, Altmann, Auffenberg, Bai, Baker, Barwick, Baum, Bay, Beatty, Bechet, Becker~Tjus, Becker, Bell, Benabderrahmane, BenZvi, Berdermann, Berghaus, Berley, Bernardini, Bernhard, Bertrand, Besson, Binder, Bindig, Bissok, Blaufuss, Blumenthal, Boersma, Bohaichuk, Bohm, Bose, Böser, Botner, Brayeur, Bretz, Brown, Bruijn, Brunner, Carson, Casey, Casier, Chirkin, Christov, Christy, Clark, Clevermann, Coenders, Cohen, Cowen, Cruz~Silva, Danninger, Daughhetee, Davis, De~Clercq, De~Ridder, Desiati, de~With, DeYoung, Díaz-Vélez, Dunkman, Eagan, Eberhardt, Eisch, Ellsworth, Euler, Evenson, Fadiran, Fazely, Fedynitch, Feintzeig, Feusels, Filimonov, Finley, Fischer-Wasels, Flis, Franckowiak, Franke, Frantzen, Fuchs, Gaisser, Gallagher, Gerhardt, Gladstone, Glüsenkamp, Goldschmidt, Golup, Gonzalez, Goodman, Góra, Grant, Groß, Gurtner, Ha, Haj~Ismail, Hallen, Hallgren, Halzen, Hanson, Heereman, Heinen, Helbing, Hellauer, Hickford,
  Hill, Hoffman, Hoffmann, Homeier, Hoshina, Huelsnitz, Hulth, Hultqvist, Hussain, Ishihara, Jacobi, Jacobsen, Jagielski, Japaridze, Jero, Jlelati, Kaminsky, Kappes, Karg, Karle, Kelley, Kiryluk, Kislat, Kläs, Klein, Köhne, Kohnen, Kolanoski, Köpke, Kopper, Kopper, Koskinen, Kowalski, Krasberg, Krings, Kroll, Kunnen, Kurahashi, Kuwabara, Labare, Landsman, Larson, Lesiak-Bzdak, Leuermann, Leute, Lünemann, Madsen, Maruyama, Mase, Matis, McNally, Meagher, Merck, Mészáros, Meures, Miarecki, Middell, Milke, Miller, Mohrmann, Montaruli, Morse, Nahnhauer, Naumann, Niederhausen, Nowicki, Nygren, Obertacke, Odrowski, Olivas, Olivo, O’Murchadha, Paul, Pepper, Pérez de~los Heros, Pfendner, Pieloth, Pinat, Pirk, Posselt, Price, Przybylski, Rädel, Rameez, Rawlins, Redl, Reimann, Resconi, Rhode, Ribordy, Richman, Riedel, Rodrigues, Rott, Ruhe, Ruzybayev, Ryckbosch, Saba, Salameh, Sander, Santander, Sarkar, Schatto, Scheel, Scheriau, Schmidt, Schmitz, Schoenen, Schöneberg, Schönwald, Schukraft, Schulte, Schulz,
  Seckel, Sestayo, Seunarine, Sheremata, Smith, Soiron, Soldin, Spiczak, Spiering, Stamatikos, Stanev, Stasik, Stezelberger, Stokstad, Stößl, Strahler, Ström, Sullivan, Taavola, Taboada, Tamburro, Ter-Antonyan, Tešić, Tilav, Toale, Toscano, Usner, van~der Drift, van Eijndhoven, Van~Overloop, van Santen, Vehring, Voge, Vraeghe, Walck, Waldenmaier, Wallraff, Wasserman, Weaver, Wellons, Wendt, Westerhoff, Whitehorn, Wiebe, Wiebusch, Williams, Wissing, Wolf, Wood, Woschnagg, Xu, Xu, Xu, Yanez, Yodh, Yoshida, Zarzhitsky, Ziemann, Zierke, Zilles, \& Zoll}]{Aartsen_2013}
Aartsen, M.~G., Abbasi, R., Abdou, Y., {et~al.} 2013, Physical Review Letters, 111

\bibitem[{{Aartsen} {et~al.}(2019){Aartsen}, {Ackermann}, {Adams}, {Aguilar}, {Ahlers}, {Ahrens}, {Alispach}, {Andeen}, {Anderson}, {Ansseau}, {Anton}, {Arg{\"u}elles}, {Auffenberg}, {Axani}, {Backes}, {Bagherpour}, {Bai}, {Barbano}, {Barwick}, {Baum}, {Baur}, {Bay}, {Beatty}, {Becker}, {Becker Tjus}, {BenZvi}, {Berley}, {Bernardini}, {Besson}, {Binder}, {Bindig}, {Blaufuss}, {Blot}, {Bohm}, {B{\"o}rner}, {B{\"o}ser}, {Botner}, {B{\"o}ttcher}, {Bourbeau}, {Bourbeau}, {Bradascio}, {Braun}, {Bretz}, {Bron}, {Brostean-Kaiser}, {Burgman}, {Buscher}, {Busse}, {Carver}, {Chen}, {Cheung}, {Chirkin}, {Clark}, {Classen}, {Collin}, {Conrad}, {Coppin}, {Correa}, {Cowen}, {Cross}, {Dave}, {de Andr{\'e}}, {De Clercq}, {DeLaunay}, {Dembinski}, {Deoskar}, {De Ridder}, {Desiati}, {de Vries}, {de Wasseige}, {de With}, {DeYoung}, {Diaz}, {D{\'\i}az-V{\'e}lez}, {Dujmovic}, {Dunkman}, {Dvorak}, {Eberhardt}, {Ehrhardt}, {Eller}, {Evenson}, {Fahey}, {Fazely}, {Felde}, {Feusels}, {Filimonov}, {Finley}, {Franckowiak}, {Friedman},
  {Fritz}, {Gaisser}, {Gallagher}, {Ganster}, {Garrappa}, {Gerhardt}, {Ghorbani}, {Glauch}, {Gl{\"u}senkamp}, {Goldschmidt}, {Gonzalez}, {Grant}, {Griffith}, {G{\"u}nder}, {G{\"u}nd{\"u}z}, {Haack}, {Hallgren}, {Halve}, {Halzen}, {Hanson}, {Hebecker}, {Heereman}, {Heix}, {Helbing}, {Hellauer}, {Henningsen}, {Hickford}, {Hignight}, {Hill}, {Hoffman}, {Hoffmann}, {Hoinka}, {Hokanson-Fasig}, {Hoshina}, {Huang}, {Huber}, {Hultqvist}, {H{\"u}nnefeld}, {Hussain}, {In}, {Iovine}, {Ishihara}, {Jacobi}, {Japaridze}, {Jeong}, {Jero}, {Jones}, {Jonske}, {Joppe}, {Kang}, {Kappes}, {Kappesser}, {Karg}, {Karl}, {Karle}, {Katz}, {Kauer}, {Kelley}, {Kheirandish}, {Kim}, {Kintscher}, {Kiryluk}, {Kittler}, {Klein}, {Koirala}, {Kolanoski}, {K{\"o}pke}, {Kopper}, {Kopper}, {Koskinen}, {Kowalski}, {Krings}, {Kr{\"u}ckl}, {Kulacz}, {Kunwar}, {Kurahashi}, {Kyriacou}, {Labare}, {Lanfranchi}, {Larson}, {Lauber}, {Lazar}, {Leonard}, {Leuermann}, {Liu}, {Lohfink}, {Lozano Mariscal}, {Lu}, {Lucarelli}, {L{\"u}nemann}, {Luszczak},
  {Madsen}, {Maggi}, {Mahn}, {Makino}, {Mallik}, {Mallot}, {Mancina}, {Mari{\c{s}}}, {Maruyama}, {Mase}, {Maunu}, {Meagher}, {Medici}, {Medina}, {Meier}, {Meighen-Berger}, {Menne}, {Merino}, {Meures}, {Miarecki}, {Micallef}, {Moment{\'e}}, {Montaruli}, {Moore}, {Morse}, {Moulai}, {Muth}, {Nagai}, {Nahnhauer}, {Nakarmi}, {Naumann}, {Neer}, {Niederhausen}, {Nowicki}, {Nygren}, {Obertacke Pollmann}, {Olivas}, {O'Murchadha}, {O'Sullivan}, {Palczewski}, {Pandya}, {Pankova}, {Park}, {Peiffer}, {P{\'e}rez de los Heros}, {Philippen}, {Pieloth}, {Pinat}, {Pizzuto}, {Plum}, {Porcelli}, {Price}, {Przybylski}, {Raab}, {Raissi}, {Rameez}, {Rauch}, {Rawlins}, {Rea}, {Reimann}, {Relethford}, {Renzi}, {Resconi}, {Rhode}, {Richman}, {Robertson}, {Rongen}, {Rott}, {Ruhe}, {Ryckbosch}, {Rysewyk}, {Safa}, {Sanchez Herrera}, {Sandrock}, {Sandroos}, {Santander}, {Sarkar}, {Sarkar}, {Satalecka}, {Schaufel}, {Schlunder}, {Schmidt}, {Schneider}, {Schneider}, {Schumacher}, {Sclafani}, {Seckel}, {Seunarine}, {Shefali}, {Silva},
  {Snihur}, {Soedingrekso}, {Soldin}, {Song}, {Spiczak}, {Spiering}, {Stachurska}, {Stamatikos}, {Stanev}, {Stasik}, {Stein}, {Stettner}, {Steuer}, {Stezelberger}, {Stokstad}, {St{\"o}{\ss}l}, {Strotjohann}, {St{\"u}rwald}, {Stuttard}, {Sullivan}, {Sutherland}, {Taboada}, {Tenholt}, {Ter-Antonyan}, {Terliuk}, {Tilav}, {Tomankova}, {T{\"o}nnis}, {Toscano}, {Tosi}, {Tselengidou}, {Tung}, {Turcati}, {Turcotte}, {Turley}, {Ty}, {Unger}, {Unland Elorrieta}, {Usner}, {Vandenbroucke}, {Van Driessche}, {van Eijk}, {van Eijndhoven}, {Vanheule}, {van Santen}, {Vraeghe}, {Walck}, {Wallace}, {Wallraff}, {Wandkowsky}, {Watson}, {Weaver}, {Weiss}, {Weldert}, {Wendt}, {Werthebach}, {Westerhoff}, {Whelan}, {Whitehorn}, {Wiebe}, {Wiebusch}, {Wille}, {Williams}, {Wills}, {Wolf}, {Wood}, {Wood}, {Woschnagg}, {Wrede}, {Xu}, {Xu}, {Xu}, {Yanez}, {Yodh}, {Yoshida}, {Yuan}, {Z{\"o}cklein}, \& {IceCube Collaboration}}]{ICECUBE2019ICRC}
{Aartsen}, M.~G., {Ackermann}, M., {Adams}, J., {et~al.} 2019, \prd, 100, 082002

\bibitem[{Abbasi {et~al.}(2025)}]{IceCube:2025ezc}
Abbasi, R. {et~al.} 2025, arXiv [\eprint[arXiv]{2502.01963}]

\bibitem[{{Abdul Halim} {et~al.}(2024){Abdul Halim}, Abreu, Aglietta, Allekotte, Cheminant, Almela, Aloisio, Alvarez-Muñiz, Yebra, Anastasi, Anchordoqui, Andrada, Andringa, Apollonio, Aramo, Ferreira, Arnone, Velázquez, Assis, Avila, Avocone, Bakalova, Barbato, Mocellin, Bellido, Berat, Bertaina, Bhatta, Bianciotto, Biermann, Binet, Bismark, Bister, Biteau, Blazek, Bleve, Blümer, Boháčová, Boncioli, Bonifazi, Arbeletche, Borodai, Brack, Orchera, Briechle, Bueno, Buitink, Buscemi, Büsken, Bwembya, Caballero-Mora, Cabana-Freire, Caccianiga, Campuzano, Caruso, Castellina, Catalani, Cataldi, Cazon, Cerda, Cermenati, Chinellato, Chudoba, Chytka, Clay, Cerutti, Colalillo, Coluccia, Conceição, Condorelli, Consolati, Conte, Convenga, dos Santos, Costa, Covault, Cristinziani, Sanchez, Dasso, Daumiller, Dawson, de~Almeida, de~Jesús, de~Jong, de~Mello~Neto, Mitri, de~Oliveira, de~Oliveira~Franco, de~Palma, de~Souza, de~Souza~de Errico, Vito, Popolo, Deligny, Denner, Deval, di~Matteo, Dobre, Dobrigkeit,
  D'Olivo, Mendes, Dorosti, dos Anjos, dos Anjos, Ebr, Ellwanger, Emam, Engel, Epicoco, Erdmann, Etchegoyen, Evoli, Falcke, Farrar, Fauth, Feldbusch, Fenu, Fernandes, Fick, Figueira, Filipčič, Fitoussi, Flaggs, Fodran, Fujii, Fuster, Galea, García, Gaudu, Gherghel-Lascu, Giaccari, Glombitza, Gobbi, Gollan, Golup, Berisso, Vitale, Gongora, González, González, Góra, Gorgi, Gottowik, Guarino, Guedes, Guido, Gülzow, Hahn, Hamal, Hampel, Hansen, Harari, Harvey, Haungs, Hebbeker, Hojvat, Hörandel, Horvath, Hrabovský, Huege, Insolia, Isar, Janardhana, Janecek, Jilek, Johnsen, Jurysek, Kampert, Keilhauer, Khakurdikar, Covilakam, Klages, Kleifges, Knapp, Köhler, Krieger, Kunka, Lago, Langner, de~Oliveira, Lema-Capeans, Letessier-Selvon, Lhenry-Yvon, Lopes, Lu, Luce, Lundquist, Payeras, Majercakova, Mandat, Manning, Mantsch, Mariani, Mariazzi, Mariş, Marsella, Martello, Martinelli, Bravo, Martins, Mathes, Matthews, Matthiae, Mayotte, Mayotte, Mazur, Medina-Tanco, Meinert, Melo, Menshikov, Merx, Michal,
  Micheletti, Miramonti, Mollerach, Montanet, Morejon, Mulrey, Mussa, Namasaka, Negi, Nellen, Nguyen, Nicora, Niechciol, Nitz, Nosek, Novotny, Nožka, Nucita, Núñez, Oliveira, Palatka, Pallotta, Panja, Parente, Paulsen, Pawlowsky, Pech, Pȩkala, Pelayo, Pelgrims, Pereira, Martins, Armand, Bertolli, Perrone, Petrera, Petrucci, Pierog, Pimenta, Platino, Pont, Pothast, Shahvar, Privitera, Prouza, Querchfeld, Rautenberg, Ravignani, Akim, Reininghaus, Reuzki, Ridky, Riehn, Risse, Rizi, de~Carvalho, Rodriguez, Rojo, Roncoroni, Rossoni, Roth, Roulet, Rovero, Ruehl, Saftoiu, Saharan, Salamida, Salazar, Salina, Gomez, Sánchez, Santos, Santos, Sarazin, Sarmento, Sato, Savina, Schäfer, Scherini, Schieler, Schimassek, Schimp, Schmidt, Scholten, Schoorlemmer, Schovánek, Schröder, Schulte, Schulz, Sciutto, Scornavacche, Sedoski, Segreto, Sehgal, Shivashankara, Sigl, Silli, Sima, Simkova, Simon, Smau, Šmída, Sommers, Soriano, Squartini, Stadelmaier, Stanič, Stasielak, Stassi, Strähnz, Straub, Suomijärvi,
  Supanitsky, Svozilikova, Szadkowski, Tairli, Tapia, Taricco, Timmermans, Tkachenko, Tobiska, Peixoto, Tomé, Torrès, Travaini, Travnicek, Tueros, Unger, Uzeiroska, Vaclavek, Vacula, Galicia, Valore, Varela, Vašíčková, Vásquez-Ramírez, Veberič, Quispe, Verzi, Vicha, Vink, Vorobiov, Watanabe, Weindl, Wiencke, Wilczyński, Wittkowski, Wundheiler, Yue, Yushkov, Zapparrata, Zas, Zavrtanik, Zavrtanik, \& collaboration}]{halim2024impact}
{Abdul Halim}, A., Abreu, P., Aglietta, M., {et~al.} 2024, Journal of Cosmology and Astroparticle Physics, 2024, 094

\bibitem[{{Abdul Halim} {et~al.}(2023){Abdul Halim}, Abreu, Aglietta, Allekotte, Cheminant, Almela, Alvarez-Muñiz, Yebra, Anastasi, Anchordoqui, Andrada, Andringa, Aramo, Ferreira, Arnone, Velázquez, Asorey, Assis, Avila, Avocone, Badescu, Bakalova, Balaceanu, Barbato, Bellido, Berat, Bertaina, Bhatta, Biermann, Binet, Bismark, Bister, Biteau, Blazek, Bleve, Blümer, Boháčová, Boncioli, Bonifazi, Arbeletche, Borodai, Brack, Bretz, Orchera, Briechle, Buchholz, Bueno, Buitink, Buscemi, Büsken, Bwembya, Caballero-Mora, Caccianiga, Caracas, Caruso, Castellina, Catalani, Cataldi, Cazon, Cerda, Chinellato, Chudoba, Chytka, Clay, Cerutti, Colalillo, Coleman, Coluccia, Conceição, Condorelli, Consolati, Conte, Contreras, Convenga, dos Santos, Covault, Cristinziani, Sanchez, Dasso, Daumiller, Dawson, de~Almeida, de~Jesús, de~Jong, de~Mello~Neto, Mitri, de~Oliveira, de~Oliveira~Franco, de~Palma, de~Souza, Vito, Popolo, Deligny, Deval, di~Matteo, Dobre, Dobrigkeit, D'Olivo, Mendes, dos Anjos, Ebr, Eman, Engel,
  Epicoco, Erdmann, Etchegoyen, Falcke, Farmer, Farrar, Fauth, Fazzini, Feldbusch, Fenu, Fick, Figueira, Filipčič, Fitoussi, Flaggs, Fodran, Fujii, Fuster, Galea, Galelli, García, Gemmeke, Gesualdi, Gherghel-Lascu, Ghia, Giaccari, Giammarchi, Glombitza, Gobbi, Gollan, Golup, Berisso, Vitale, Gongora, González, González, Goos, Góra, Gorgi, Gottowik, Grubb, Guarino, Guedes, Guido, Hahn, Hamal, Hampel, Hansen, Harari, Harvey, Haungs, Hebbeker, Heck, Hojvat, Hörandel, Horvath, Hrabovský, Huege, Insolia, Isar, Janecek, Johnsen, Jurysek, Kääpä, Kampert, Keilhauer, Khakurdikar, Covilakam, Klages, Kleifges, Kleinfeller, Knapp, Kunka, Lago, Langner, de~Oliveira, Lenok, Letessier-Selvon, Lhenry-Yvon, Presti, Lopes, López, Lu, Luce, Lundquist, Payeras, Majercakova, Mandat, Manning, Manshanden, Mantsch, Marafico, Mariani, Mariazzi, Mariş, Marsella, Martello, Martinelli, Bravo, Martins, Mastrodicasa, Mathes, Matthews, Matthiae, Mayotte, Mayotte, Mazur, Medina-Tanco, Meinert, Melo, Menshikov, Michal,
  Micheletti, Miramonti, Mollerach, Montanet, Morejon, Morello, Müller, Mulrey, Mussa, Muzio, Namasaka, Nasr-Esfahani, Nellen, Nicora, Niculescu-Oglinzanu, Niechciol, Nitz, Norwood, Nosek, Novotny, Nožka, Nucita, Núñez, Oliveira, Palatka, Pallotta, Parente, Parra, Pawlowsky, Pech, Pȩkala, Pelayo, Martins, Armand, Bertolli, Perrone, Petrera, Petrucci, Pierog, Pimenta, Platino, Pont, Pothast, Shavar, Privitera, Prouza, Puyleart, Querchfeld, Rautenberg, Ravignani, Reininghaus, Ridky, Riehn, Risse, Rizi, de~Carvalho, Rojo, Roncoroni, Rossoni, Roth, Roulet, Rovero, Ruehl, Saftoiu, Saharan, Salamida, Salazar, Salina, Gomez, Sánchez, Santos, Santos, Sarazin, Sarmento, Sato, Savina, Schäfer, Scherini, Schieler, Schimassek, Schimp, Schlüter, Schmidt, Scholten, Schoorlemmer, Schovánek, Schröder, Schulte, Schulz, Sciutto, Scornavacche, Segreto, Sehgal, Shivashankara, Sigl, Silli, Sima, Smau, Šmída, Sommers, Soriano, Squartini, Stadelmaier, Stanca, Stanič, Stasielak, Stassi, Straub, Streich, Suárez-Durán,
  Suomijärvi, Supanitsky, Szadkowski, Tapia, Taricco, Timmermans, Tkachenko, Tobiska, Peixoto, Tomé, Torrès, Travaini, Travnicek, Trimarelli, Tueros, Ulrich, Unger, Vaclavek, Vacula, Galicia, Valore, Varela, Vásquez-Ramírez, Veberič, Ventura, Quispe, Verzi, Vicha, Vink, Vorobiov, Watanabe, Watson, Weindl, Wiencke, Wilczyński, Wittkowski, Wundheiler, Yushkov, Zapparrata, Zas, Zavrtanik, Zavrtanik, \& collaboration}]{AbdulHalim_2023}
{Abdul Halim}, A., Abreu, P., Aglietta, M., {et~al.} 2023, Journal of Cosmology and Astroparticle Physics, 2023, 024

\bibitem[{Abreu {et~al.}(2021)Abreu, Aglietta, Albury, Allekotte, Almela, Alvarez-Mu{\~n}iz, Alves~Batista, Anastasi, Anchordoqui, Andrada, {et~al.}}]{abreu2021energy}
Abreu, P., Aglietta, M., Albury, J.~M., {et~al.} 2021, The European Physical Journal C, 81, 1

\bibitem[{Ajello {et~al.}(2009)Ajello, Costamante, Sambruna, Gehrels, Chiang, Rau, Escala, Greiner, Tueller, Wall, \& Mushotzky}]{Ajello_2009}
Ajello, M., Costamante, L., Sambruna, R.~M., {et~al.} 2009, The Astrophysical Journal, 699, 603

\bibitem[{Ajello {et~al.}(2012)Ajello, Shaw, Romani, Dermer, Costamante, King, Max-Moerbeck, Readhead, Reimer, Richards, \& Stevenson}]{Ajello_2012}
Ajello, M., Shaw, M.~S., Romani, R.~W., {et~al.} 2012, The Astrophysical Journal, 751, 108

\bibitem[{Aloisio \& Berezinsky(2004)}]{Aloisio_2004}
Aloisio, R. \& Berezinsky, V. 2004, The Astrophysical Journal, 612, 900

\bibitem[{Aloisio {et~al.}(2017)Aloisio, Boncioli, di~Matteo, Grillo, Petrera, \& Salamida}]{SimProp2017}
Aloisio, R., Boncioli, D., di~Matteo, A., {et~al.} 2017, Journal of Cosmology and Astroparticle Physics, 2017, 009

\bibitem[{Amaral {et~al.}(2021)Amaral, Vernstrom, \& Gaensler}]{Amaral2021}
Amaral, A.~D., Vernstrom, T., \& Gaensler, B.~M. 2021, Monthly Notices of the Royal Astronomical Society, 503, 2913

\bibitem[{Amato \& Blasi(2009)}]{BlasiAmatoKinetic}
Amato, E. \& Blasi, P. 2009, Monthly Notices of the Royal Astronomical Society, 392, 1591

\bibitem[{{Arons}(2003)}]{Arons2003}
{Arons}, J. 2003, \apj, 589, 871

\bibitem[{{Asano} \& {M{\'e}sz{\'a}ros}(2016)}]{Asano2016}
{Asano}, K. \& {M{\'e}sz{\'a}ros}, P. 2016, \prd, 94, 023005

\bibitem[{{Bell}(2004)}]{Bell2004mnras}
{Bell}, A.~R. 2004, \mnras, 353, 550

\bibitem[{Berezinsky {et~al.}(2006)Berezinsky, Gazizov, \& Grigorieva}]{Berezinsky_Gazizov_prd}
Berezinsky, V., Gazizov, A., \& Grigorieva, S. 2006, Phys. Rev. D, 74, 043005

\bibitem[{Berezinsky {et~al.}(1997)Berezinsky, Blasi, \& Ptuskin}]{Berezinsky_1997}
Berezinsky, V.~S., Blasi, P., \& Ptuskin, V.~S. 1997, The Astrophysical Journal, 487, 529

\bibitem[{Bergman(2006)}]{Douglas_Bergman_2006}
Bergman, D. 2006, Journal of Physics: Conference Series, 47, 154

\bibitem[{Blandford \& Eichler(1987)}]{BLANDFORD19871}
Blandford, R. \& Eichler, D. 1987, Physics Reports, 154, 1

\bibitem[{{Blandford} \& {Ostriker}(1978)}]{Blandford1978ApJ}
{Blandford}, R.~D. \& {Ostriker}, J.~P. 1978, \apjl, 221, L29

\bibitem[{Blasi \& Amato(2019)}]{blasi2019prl}
Blasi, P. \& Amato, E. 2019, Phys. Rev. Lett., 122, 051101

\bibitem[{{Blasi} {et~al.}(2015){Blasi}, {Amato}, \& {D'Angelo}}]{Blasi2015prl}
{Blasi}, P., {Amato}, E., \& {D'Angelo}, M. 2015, \prl, 115, 121101

\bibitem[{{Blasi} {et~al.}(2000){Blasi}, {Epstein}, \& {Olinto}}]{Blasi2000}
{Blasi}, P., {Epstein}, R.~I., \& {Olinto}, A.~V. 2000, \apjl, 533, L123

\bibitem[{Burlon {et~al.}(2011)Burlon, Ajello, Greiner, Comastri, Merloni, \& Gehrels}]{Burlon_2011}
Burlon, D., Ajello, M., Greiner, J., {et~al.} 2011, The Astrophysical Journal, 728, 58

\bibitem[{Carretti {et~al.}(2022)Carretti, O’Sullivan, Vacca, Vazza, Gheller, Vernstrom, \& Bonafede}]{Carretti2022}
Carretti, E., O’Sullivan, S.~P., Vacca, V., {et~al.} 2022, Monthly Notices of the Royal Astronomical Society, 518, 2273

\bibitem[{{Ensslin} {et~al.}(1997){Ensslin}, {Biermann}, {Kronberg}, \& {Wu}}]{Ensslin1997}
{Ensslin}, T.~A., {Biermann}, P.~L., {Kronberg}, P.~P., \& {Wu}, X.-P. 1997, \apj, 477, 560

\bibitem[{{Fotopoulou, S.} {et~al.}(2016){Fotopoulou, S.}, {Buchner, J.}, {Georgantopoulos, I.}, {Hasinger, G.}, {Salvato, M.}, {Georgakakis, A.}, {Cappelluti, N.}, {Ranalli, P.}, {Hsu, L. T.}, {Brusa, M.}, {Comastri, A.}, {Miyaji, T.}, {Nandra, K.}, {Aird, J.}, \& {Paltani, S.}}]{Fotopoulou2016}
{Fotopoulou, S.}, {Buchner, J.}, {Georgantopoulos, I.}, {et~al.} 2016, A\&A, 587, A142

\bibitem[{Gargaté {et~al.}(2010)Gargaté, Fonseca, Niemiec, Pohl, Bingham, \& Silva}]{gargate2010aas}
Gargaté, L., Fonseca, R.~A., Niemiec, J., {et~al.} 2010, The Astrophysical Journal Letters, 711, L127

\bibitem[{Kang(2023)}]{Kaskade2023ICRC}
Kang, D. 2023, PoS, ICRC2023, 307

\bibitem[{Kelner \& Aharonian(2010)}]{Kelner2008prd}
Kelner, S.~R. \& Aharonian, F.~A. 2010, Phys. Rev. D, 82, 099901

\bibitem[{Kopper(2017)}]{Kopper2017IceCubeLims}
Kopper, C. 2017, PoS, ICRC2017, 981

\bibitem[{{Kotera} {et~al.}(2015){Kotera}, {Amato}, \& {Blasi}}]{Kotera2015}
{Kotera}, K., {Amato}, E., \& {Blasi}, P. 2015, \jcap, 2015, 026

\bibitem[{Mollerach \& Roulet(2013)}]{MollerachJCAP}
Mollerach, S. \& Roulet, E. 2013, Journal of Cosmology and Astroparticle Physics, 2013, 013

\bibitem[{{M{\"u}cke} {et~al.}(2000){M{\"u}cke}, {Engel}, {Rachen}, {Protheroe}, \& {Stanev}}]{Sophia_code}
{M{\"u}cke}, A., {Engel}, R., {Rachen}, J.~P., {Protheroe}, R.~J., \& {Stanev}, T. 2000, Computer Physics Communications, 124, 290

\bibitem[{Muzio {et~al.}(2024)Muzio, Anchordoqui, \& Unger}]{muzio2024peterscycleendcosmic}
Muzio, M.~S., Anchordoqui, L.~A., \& Unger, M. 2024, A Peters cycle at the end of the cosmic ray spectrum?

\bibitem[{Muzio {et~al.}(2022)Muzio, Farrar, \& Unger}]{Muzio2022PRD}
Muzio, M.~S., Farrar, G.~R., \& Unger, M. 2022, Phys. Rev. D, 105, 023022

\bibitem[{O’Sullivan {et~al.}(2020)O’Sullivan, Brüggen, Vazza, Carretti, Locatelli, Stuardi, Vacca, Vernstrom, Heald, Horellou, Shimwell, Hardcastle, Tasse, \& Röttgering}]{OSullivan2020}
O’Sullivan, S.~P., Brüggen, M., Vazza, F., {et~al.} 2020, Monthly Notices of the Royal Astronomical Society, 495, 2607

\bibitem[{Paccagnella {et~al.}(2017)Paccagnella, Vulcani, Poggianti, Fritz, Fasano, Moretti, Jaffé, Biviano, Gullieuszik, Bettoni, Cava, Couch, \& D’Onofrio}]{Paccagnella_2017}
Paccagnella, A., Vulcani, B., Poggianti, B.~M., {et~al.} 2017, The Astrophysical Journal, 838, 148

\bibitem[{Pierog {et~al.}(2015)Pierog, Karpenko, \& Katzy}]{pierog2015epos}
Pierog, T., Karpenko, I., \& Katzy, J. M. e.~a. 2015, Physical Review C, 92, 034906

\bibitem[{Pimbblet {et~al.}(2012)Pimbblet, Shabala, Haines, Fraser-McKelvie, \& Floyd}]{Pimbblet2012}
Pimbblet, K.~A., Shabala, S.~S., Haines, C.~P., Fraser-McKelvie, A., \& Floyd, D. J.~E. 2012, Monthly Notices of the Royal Astronomical Society, 429, 1827

\bibitem[{{Planck Collaboration} {et~al.}(2016){Planck Collaboration}, {Ade, P. A. R.}, {Aghanim, N.}, {Arnaud, M.}, {Ashdown, M.}, {Aumont, J.}, {Baccigalupi, C.}, {Banday, A. J.}, {Barreiro, R. B.}, {Bartlett, J. G.}, {Bartolo, N.}, {Battaner, E.}, {Battye, R.}, {Benabed, K.}, {Benoît, A.}, {Benoit-Lévy, A.}, {Bernard, J.-P.}, {Bersanelli, M.}, {Bielewicz, P.}, {Bock, J. J.}, {Bonaldi, A.}, {Bonavera, L.}, {Bond, J. R.}, {Borrill, J.}, {Bouchet, F. R.}, {Boulanger, F.}, {Bucher, M.}, {Burigana, C.}, {Butler, R. C.}, {Calabrese, E.}, {Cardoso, J.-F.}, {Catalano, A.}, {Challinor, A.}, {Chamballu, A.}, {Chary, R.-R.}, {Chiang, H. C.}, {Chluba, J.}, {Christensen, P. R.}, {Church, S.}, {Clements, D. L.}, {Colombi, S.}, {Colombo, L. P. L.}, {Combet, C.}, {Coulais, A.}, {Crill, B. P.}, {Curto, A.}, {Cuttaia, F.}, {Danese, L.}, {Davies, R. D.}, {Davis, R. J.}, {de Bernardis, P.}, {de Rosa, A.}, {de Zotti, G.}, {Delabrouille, J.}, {Désert, F.-X.}, {Di Valentino, E.}, {Dickinson, C.}, {Diego, J. M.}, {Dolag, K.},
  {Dole, H.}, {Donzelli, S.}, {Doré, O.}, {Douspis, M.}, {Ducout, A.}, {Dunkley, J.}, {Dupac, X.}, {Efstathiou, G.}, {Elsner, F.}, {Enßlin, T. A.}, {Eriksen, H. K.}, {Farhang, M.}, {Fergusson, J.}, {Finelli, F.}, {Forni, O.}, {Frailis, M.}, {Fraisse, A. A.}, {Franceschi, E.}, {Frejsel, A.}, {Galeotta, S.}, {Galli, S.}, {Ganga, K.}, {Gauthier, C.}, {Gerbino, M.}, {Ghosh, T.}, {Giard, M.}, {Giraud-Héraud, Y.}, {Giusarma, E.}, {Gjerløw, E.}, {González-Nuevo, J.}, {Górski, K. M.}, {Gratton, S.}, {Gregorio, A.}, {Gruppuso, A.}, {Gudmundsson, J. E.}, {Hamann, J.}, {Hansen, F. K.}, {Hanson, D.}, {Harrison, D. L.}, {Helou, G.}, {Henrot-Versillé, S.}, {Hernández-Monteagudo, C.}, {Herranz, D.}, {Hildebrandt, S. R.}, {Hivon, E.}, {Hobson, M.}, {Holmes, W. A.}, {Hornstrup, A.}, {Hovest, W.}, {Huang, Z.}, {Huffenberger, K. M.}, {Hurier, G.}, {Jaffe, A. H.}, {Jaffe, T. R.}, {Jones, W. C.}, {Juvela, M.}, {Keihänen, E.}, {Keskitalo, R.}, {Kisner, T. S.}, {Kneissl, R.}, {Knoche, J.}, {Knox, L.}, {Kunz, M.},
  {Kurki-Suonio, H.}, {Lagache, G.}, {Lähteenmäki, A.}, {Lamarre, J.-M.}, {Lasenby, A.}, {Lattanzi, M.}, {Lawrence, C. R.}, {Leahy, J. P.}, {Leonardi, R.}, {Lesgourgues, J.}, {Levrier, F.}, {Lewis, A.}, {Liguori, M.}, {Lilje, P. B.}, {Linden-Vørnle, M.}, {López-Caniego, M.}, {Lubin, P. M.}, {Macías-Pérez, J. F.}, {Maggio, G.}, {Maino, D.}, {Mandolesi, N.}, {Mangilli, A.}, {Marchini, A.}, {Maris, M.}, {Martin, P. G.}, {Martinelli, M.}, {Martínez-González, E.}, {Masi, S.}, {Matarrese, S.}, {McGehee, P.}, {Meinhold, P. R.}, {Melchiorri, A.}, {Melin, J.-B.}, {Mendes, L.}, {Mennella, A.}, {Migliaccio, M.}, {Millea, M.}, {Mitra, S.}, {Miville-Deschênes, M.-A.}, {Moneti, A.}, {Montier, L.}, {Morgante, G.}, {Mortlock, D.}, {Moss, A.}, {Munshi, D.}, {Murphy, J. A.}, {Naselsky, P.}, {Nati, F.}, {Natoli, P.}, {Netterfield, C. B.}, {Nørgaard-Nielsen, H. U.}, {Noviello, F.}, {Novikov, D.}, {Novikov, I.}, {Oxborrow, C. A.}, {Paci, F.}, {Pagano, L.}, {Pajot, F.}, {Paladini, R.}, {Paoletti, D.}, {Partridge, B.},
  {Pasian, F.}, {Patanchon, G.}, {Pearson, T. J.}, {Perdereau, O.}, {Perotto, L.}, {Perrotta, F.}, {Pettorino, V.}, {Piacentini, F.}, {Piat, M.}, {Pierpaoli, E.}, {Pietrobon, D.}, {Plaszczynski, S.}, {Pointecouteau, E.}, {Polenta, G.}, {Popa, L.}, {Pratt, G. W.}, {Prézeau, G.}, {Prunet, S.}, {Puget, J.-L.}, {Rachen, J. P.}, {Reach, W. T.}, {Rebolo, R.}, {Reinecke, M.}, {Remazeilles, M.}, {Renault, C.}, {Renzi, A.}, {Ristorcelli, I.}, {Rocha, G.}, {Rosset, C.}, {Rossetti, M.}, {Roudier, G.}, {Rouillé d’Orfeuil, B.}, {Rowan-Robinson, M.}, {Rubiño-Martín, J. A.}, {Rusholme, B.}, {Said, N.}, {Salvatelli, V.}, {Salvati, L.}, {Sandri, M.}, {Santos, D.}, {Savelainen, M.}, {Savini, G.}, {Scott, D.}, {Seiffert, M. D.}, {Serra, P.}, {Shellard, E. P. S.}, {Spencer, L. D.}, {Spinelli, M.}, {Stolyarov, V.}, {Stompor, R.}, {Sudiwala, R.}, {Sunyaev, R.}, {Sutton, D.}, {Suur-Uski, A.-S.}, {Sygnet, J.-F.}, {Tauber, J. A.}, {Terenzi, L.}, {Toffolatti, L.}, {Tomasi, M.}, {Tristram, M.}, {Trombetti, T.}, {Tucci, M.},
  {Tuovinen, J.}, {Türler, M.}, {Umana, G.}, {Valenziano, L.}, {Valiviita, J.}, {Van Tent, F.}, {Vielva, P.}, {Villa, F.}, {Wade, L. A.}, {Wandelt, B. D.}, {Wehus, I. K.}, {White, M.}, {White, S. D. M.}, {Wilkinson, A.}, {Yvon, D.}, {Zacchei, A.}, \& {Zonca, A.}}]{Planck2015}
{Planck Collaboration}, {Ade, P. A. R.}, {Aghanim, N.}, {et~al.} 2016, A\&A, 594, A13

\bibitem[{Qu {et~al.}(2019)Qu, Zeng, \& Yan}]{qu2019}
Qu, Y., Zeng, H., \& Yan, D. 2019, Monthly Notices of the Royal Astronomical Society, 490, 758

\bibitem[{Riquelme \& Spitkovsky(2009)}]{Riquelmeapj}
Riquelme, M.~A. \& Spitkovsky, A. 2009, The Astrophysical Journal, 694, 626

\bibitem[{Saldana-Lopez {et~al.}(2021)Saldana-Lopez, Domínguez, Pérez-González, Finke, Ajello, Primack, Paliya, \& Desai}]{saldana2021ebl}
Saldana-Lopez, A., Domínguez, A., Pérez-González, P.~G., {et~al.} 2021, Monthly Notices of the Royal Astronomical Society, 507, 5144

\bibitem[{{Schroer} {et~al.}(2021){Schroer}, {Pezzi}, {Caprioli}, {Haggerty}, \& {Blasi}}]{Schroer2021}
{Schroer}, B., {Pezzi}, O., {Caprioli}, D., {Haggerty}, C., \& {Blasi}, P. 2021, \apjl, 914, L13

\bibitem[{{Schroer} {et~al.}(2022){Schroer}, {Pezzi}, {Caprioli}, {Haggerty}, \& {Blasi}}]{Schroer2022}
{Schroer}, B., {Pezzi}, O., {Caprioli}, D., {Haggerty}, C.~C., \& {Blasi}, P. 2022, \mnras, 512, 233

\bibitem[{Shinozaki \& Teshima(2004)}]{SHINOZAKI200418}
Shinozaki, K. \& Teshima, M. 2004, Nuclear Physics B - Proceedings Supplements, 136, 18

\bibitem[{{Sironi} {et~al.}(2015){Sironi}, {Keshet}, \& {Lemoine}}]{Sironi2015}
{Sironi}, L., {Keshet}, U., \& {Lemoine}, M. 2015, \ssr, 191, 519

\bibitem[{{Subedi} {et~al.}(2017){Subedi}, {Sonsrettee}, {Blasi}, {Ruffolo}, {Matthaeus}, {Montgomery}, {Chuychai}, {Dmitruk}, {Wan}, {Parashar}, \& {Chhiber}}]{Subedi2017}
{Subedi}, P., {Sonsrettee}, W., {Blasi}, P., {et~al.} 2017, \apj, 837, 140

\bibitem[{Unger {et~al.}(2015)Unger, Farrar, \& Anchordoqui}]{UngerPRD}
Unger, M., Farrar, G.~R., \& Anchordoqui, L.~A. 2015, Phys. Rev. D, 92, 123001

\bibitem[{Vernstrom {et~al.}(2021)Vernstrom, Heald, Vazza, Galvin, West, Locatelli, Fornengo, \& Pinetti}]{VernstromMNRAS}
Vernstrom, T., Heald, G., Vazza, F., {et~al.} 2021, Monthly Notices of the Royal Astronomical Society, 505, 4178

\bibitem[{{Zhang} {et~al.}(2023){Zhang}, {Sironi}, {Giannios}, \& {Petropoulou}}]{Sironi2023}
{Zhang}, H., {Sironi}, L., {Giannios}, D., \& {Petropoulou}, M. 2023, \apjl, 956, L36

\bibitem[{Zweibel \& Everett(2010)}]{zweibel2010aas}
Zweibel, E.~G. \& Everett, J.~E. 2010, The Astrophysical Journal, 709, 1412

\end{thebibliography}

\end{document}